\documentclass[%
 reprint,
 amsmath,amssymb,
prb,
]{revtex4-2}

\usepackage{lipsum}
\usepackage{bm}
\usepackage{amssymb}
\usepackage{amstext}
\usepackage{amsmath}
\usepackage{mathrsfs}
\usepackage{graphicx}
\usepackage{color}
\usepackage{amsthm}
\usepackage{floatrow}
\usepackage{array}

\usepackage{listings}
\usepackage{booktabs}
\usepackage{multirow}
\usepackage{hyperref}

\begin{document}
\preprint{APS/123-QED}
\title{Network comparison via encoding, decoding, and causality}
\thanks{Correspondence of this paper should be addressed to P.S (peisun@tsinghua.edu.cn)}%

 \author{Yang Tian}
\email{tiany20@mails.tsinghua.edu.cn \& tyanyang04@gmail.com}
 \altaffiliation[]{Department of Psychology \& Tsinghua Laboratory of Brain and Intelligence, Tsinghua University, Beijing, 100084, China.}
 \altaffiliation[Also at ]{Laboratory of Advanced Computing and Storage, Central Research Institute, 2012 Laboratories, Huawei Technologies Co. Ltd., Beijing, 100084, China.}

   \author{Hedong Hou}
\email{hedong.hou@etu.u-paris.fr}
 \altaffiliation[]{Institut de Math{\'e}matiques d'Orsay, 14 Rue Joliot-Curie, 91190 Gif-sur-Yvette, France.}

 \author{Guangzheng Xu}
\email{gzxu98@gmail.com}
 \altaffiliation[]{Department of Computer Science, University College London, London, WC1E 6AE, UK.}

  \author{Ziyang Zhang}
\email{zhangziyang11@huawei.com}
 \altaffiliation[]{Laboratory of Advanced Computing and Storage, Central Research Institute, 2012 Laboratories, Huawei Technologies Co. Ltd., Beijing, 100084, China.}
 
\author{Pei Sun}%
 \email{peisun@tsinghua.edu.cn}
 \altaffiliation[]{Department of Psychology \& Tsinghua Laboratory of Brain and Intelligence, Tsinghua University, Beijing, 100084, China.}
 



\begin{abstract}
 Quantifying the relations (e.g., similarity) between complex networks paves the way for studying the latent information shared across networks. However, fundamental relation metrics are not well-defined between networks. As a compromise, prevalent techniques measure network relations in data-driven manners, which are inapplicable to analytic derivations in physics. To resolve this issue, we present a theory for obtaining an optimal characterization of network topological properties. We show that a network can be fully represented by a Gaussian variable defined by a function of the Laplacian, which simultaneously satisfies network-topology-dependent smoothness and maximum entropy properties. Based on it, we can analytically measure diverse relations between complex networks. As illustrations, we define encoding (e.g., information divergence and mutual information), decoding (e.g., Fisher information), and causality (e.g., Granger causality and conditional mutual information) between networks. We validate our framework on representative networks (e.g., random networks, protein structures, and chemical compounds) to demonstrate that a series of science and engineering challenges (e.g., network evolution, embedding, and query) can be tackled from a new perspective. An implementation of our theory is released as a multi-platform toolbox.
   
\end{abstract}

\maketitle
\section{Introduction}\label{Sec1}
Complex networks are universal across different disciplines \cite{boccaletti2006complex}. Important topics in physics (e.g., quantum system characterization \cite{biamonte2019complex,mulken2011continuous,bianconi2002quantum} and non-equilibrium dynamics analysis \cite{roudi2011mean,dorogovtsev2008critical,sanchez2002nonequilibrium}), biology (e.g., brain \cite{zhou2006hierarchical,bullmore2009complex,li2011correlation,bassett2017network,sporns2000theoretical}, metabolic \cite{jeong2000large,wagner2001small,tanaka2005scale}, and protein \cite{jeong2001lethality,yook2004functional,wagner2001yeast} networks analysis), computer science (e.g., internet analysis \cite{vazquez2002large,pastor2004evolution,kahng2002robustness} and information tracking \cite{de2004fluctuations,de2004separating}), and social science (e.g., scientific community \cite{newman2001structure,tsallis2000citations,menczer2004correlated} and opinion formation \cite{sznajd2000opinion,pluchino2005changing,stauffer2004simulation} modelling) all benefit from complex networks studies \cite{boccaletti2006complex}.

However, critical challenges to network theories persistently arise due to the increasingly diverse application needs \cite{boccaletti2006complex}. Among these challenges, a fundamental yet intractable one concerns how to quantify the relations (e.g., similarity) between different complex networks \cite{soundarajan2014guide}. To date, mainstream metrics of network relations are developed in the contexts of network embedding, matching, and kernel, three computation-oriented and data-driven perspectives \cite{soundarajan2014guide,narayanan2017graph2vec}.  Comprehensive reviews of these three perspectives can be found in Refs. \cite{soundarajan2014guide,attar2017classification}, Refs. \cite{conte2004thirty,foggia2014graph,yan2016short}, and Refs. \cite{gao2010survey,borgwardt2020graph}, respectively. In general, \emph{embedding-based} approaches follow preset rules to embed networks into low-dimensional metric spaces and calculate distances between networks \cite{soundarajan2014guide,attar2017classification,narayanan2017graph2vec}. These approaches critically depend on embedding rule designs and may lack universal generalization capacities \cite{soundarajan2014guide}. \emph{Matching-based} approaches, such as exact \cite{de2003large} and inexact \cite{gao2010survey} matching, search for node mappings between networks to realize optimal matching and measure similarity \cite{foggia2014graph}. These approaches essentially deal with a kind of quadratic programming problems \cite{conte2004thirty,cour2006balanced,jiang2017graph} that are NP-hard \cite{conte2004thirty} and require relaxations of problem constraints to find approximate solutions \cite{enqvist2009optimal,leordeanu2009integer,zaslavskiy2008path}. \emph{Kernel-based} methods evaluate the similarity between networks by decomposing them into series of atomic substructures (e.g., graphlets \cite{shervashidze2009efficient}, random walks \cite{kashima2004kernels}, shortest paths \cite{borgwardt2005shortest}, and cycles \cite{horvath2004cyclic}) and measuring kernel value among these substructures (i.e., counting the number of shared substructures) \cite{gao2010survey,borgwardt2020graph}. While these substructures can reflect network topology properties, they are essentially handcrafted \cite{narayanan2017graph2vec}, i.e., extracted by certain manually defined functions, and may imply extremely high-dimensional, sparse, and non-smooth representations with poor generalization capacities \cite{yanardag2015deep}. In sum, while embedding-, matching-, and kernel-based approaches have been extensively tested on empirical data (e.g., neural data \cite{mheich2020brain,tomlinson2022regression,borgwardt2020graph,abbas2020geff}), they are inevitably limited by computational complexity (e.g., matching-based) or the dependence on empirical choice of network features (e.g., embedding-based) and kernel functions (e.g., kernel-based) \cite{mheich2020brain}. Even in cases where these methods are computationally optimal, they may still be unsatisfactory because they do not derive intrinsic relations between complex networks analytically and universally. 

Analytic metrics of network relations are indispensable for studying the physics of complex networks \cite{cimini2019statistical} but remain unknown. Certainly, one can simplify the distance between networks as the Kolmogorov–Smirnov statistic between their degree distributions (e.g., see discussions in Ref. \cite{tomlinson2022regression}) or the norm distance between their adjacency matrices (or Laplacian operators). However, these approaches either require that two networks share the same size or achieve non-ideal performance in network comparison (e.g., see results in Ref. \cite{petric2019got}).

To suggest a way to define analytic network relations, we develop an optimal characterization of complex networks that simultaneously ensures smoothness (for better reflection of network topology \cite{petric2019got,shuman2013emerging,sandryhaila2013discrete,sandryhaila2014discrete}) and maximum entropy (for better support of information-theoretical analysis \cite{cover1999elements}) properties in Secs. \ref{Sec2}-\ref{Sec3}. The derived characterizations turn out to be specific Gaussian variables defined by the functions of the Laplacian operators of complex networks. Based on this result, we can define analytic relation metrics (e.g., information divergence \cite{cover1999elements}, mutual information \cite{cover1999elements}, Fisher information \cite{cover1999elements}, and causality \cite{hlavavckova2007causality}) between networks in Sec. \ref{Sec4} and explore their generalization in Sec. \ref{Sec5}. In Secs. \ref{Sec6}-\ref{Sec7}, we demonstrate our approach on representative complex networks to realize network comparison by encoding, decoding, and causal analyses. A toolbox is provided in https://github.com/doloMing/Encoding-decoding-and-causality-between-complex-networks \cite{yang2022toolbox}.

\section{Question definition}\label{Sec2}
To suggest a potential direction, we consider:
\begin{itemize}
    \item[(I) ] How to develop an analytic and universal characterization of network topology that is free of subjective selection of topological properties and computational optimization problems?
    \item[(II) ] How to enable the characterization derived in question (I) to define analytic metrics of network relation without further constraints?
\end{itemize}

As we have mentioned in Sec. \ref{Sec1}, a simple solution of question (II), such as the Kolmogorov–Smirnov statistic between degree distributions, can not fully satisfy the needs of application. Therefore, we consider more informative metrics, including information divergence \cite{cover1999elements}, mutual information \cite{cover1999elements}, Fisher information \cite{cover1999elements}, and causality \cite{hlavavckova2007causality}, as potential candidates. These metrics, at least in our case, require a probabilistic solution (e.g., define a network as a random variable) of question (I). 

This idea inspires us to consider a mapping $\phi:V\rightarrow\Omega$ between a network $\mathcal{G}\left(V,E\right)$ without self-loops and a probability space $\left(\Omega,\mathcal{F},\mathcal{P}\right)$ with $\Omega=\mathbb{R}$. Here $V$ and $E$ denote the node and edge sets of network $\mathcal{G}$, respectively. Function $\phi$ defines a random variable $\mathcal{X}_{\phi}=\left(X_{\phi}\left(1\right),\ldots,X_{\phi}\left(n\right)\right)$ distributed on node set $V$, where $X_{\phi}\left(i\right)=\phi\left(v_{i}\right)$ and $n=\vert V\vert$ (see Fig. 1 for illustrations).

To properly reflect the network topology of $\mathcal{G}$ by $\mathcal{X}_{\phi}$, we need to consider the \emph{smoothness} of mapping $\phi$ on $\mathcal{G}$ measured by the \emph{discrete $2$-Dirichlet form} of $\phi$ \cite{shuman2013emerging}
\begin{align}
    \mathcal{S}\left(\phi\right)=\frac{1}{2}\sum_{v_{i}\in V}\sum_{\left(v_{i},v_{j}\right)\in E}\left(\frac{\partial\phi}{\partial \left(v_{i},v_{j}\right)}\bigg\vert_{v_{i}}\right)^{2},\label{EQ1}
\end{align}
where $\left(v_{i},v_{j}\right)$ denotes the edge between nodes $v_{i}$ and $v_{j}$. In Eq. (\ref{EQ1}), the edge derivative of $\phi$ with respect to edge $\left(v_{i},v_{j}\right)$ at node $v_{i}$ is defined as \cite{shuman2013emerging}
\begin{align}
    \frac{\partial\phi}{\partial \left(v_{i},v_{j}\right)}\bigg\vert_{v_{i}}=\sqrt{W_{i,j}}\left( X_{\phi}\left(j\right)-X_{\phi}\left(i\right)\right),\label{EQ2}
\end{align}
where $W$ is the non-negative weighted adjacent matrix of $\mathcal{G}$. The smaller $\mathcal{S}\left(\phi\right)$ in Eq. (\ref{EQ1}) is, the smoother $\phi$ is on $\mathcal{G}$. To understand why the smoothness of $\phi$ matters in defining $\mathcal{X}_{\phi}$ to reflect the network topology of $\mathcal{G}$, we need to consider the combinatorial Laplacian $L$ \cite{biyikoglu2007laplacian} of $\mathcal{G}$
\begin{align}
    L=\operatorname{diag}\left(\left[\operatorname{deg}\left(v_{1}\right),\ldots,\operatorname{deg}\left(v_{n}\right)\right]\right)-W,\label{EQ3}
\end{align}
where $\operatorname{diag}\left(\cdot\right)$ generates a diagonal matrix and operator $\operatorname{deg}\left(\cdot\right)$ measures node degree. Note that $\operatorname{deg}\left(v_{i}\right)=\sum_{j=1}^{n}W_{ij}$ in a weighted network. Laplacian $L$ captures key topology information of network $\mathcal{G}$ (e.g., connected components, random walks on network, and latent Laplace-Beltrami operator \cite{chung1997spectral}), which has been extensively used in spectral graph theory \cite{chung1997spectral} and graph signal theory \cite{shuman2013emerging}. The first connection between Laplacian $L$ and the smoothness of $\phi$ is well-known \cite{shuman2013emerging}
\begin{align}
     \mathcal{S}\left(\phi\right)={\mathcal{X}_{\phi}}^{T}L \mathcal{X}_{\phi},\label{EQ4}
\end{align}
which suggests that the smoothness of $\phi$ can be defined by Laplacian $L$ (see Fig. 1). The second connection is derived from the Courant-Fischer theorem \cite{horn2012matrix}, which suggests that the smoothness of $\phi$ is related to the eigenvectors and eigenvalues of Laplacian $L$. Eigenvectors with smaller eigenvalues imply a smoother $\phi$ \cite{shuman2013emerging}. Taken together, the smoothness of $\phi$ matters in our analysis because it is closely related to the topology information conveyed by Laplacian $L$. To enable variable $\mathcal{X}_{\phi}$ to represent network $\mathcal{G}$, we expect that the smoothness of $\phi$ is completely determined by the topology properties of $\mathcal{G}$.

 \begin{figure}[b!]
\includegraphics[width=1\columnwidth]{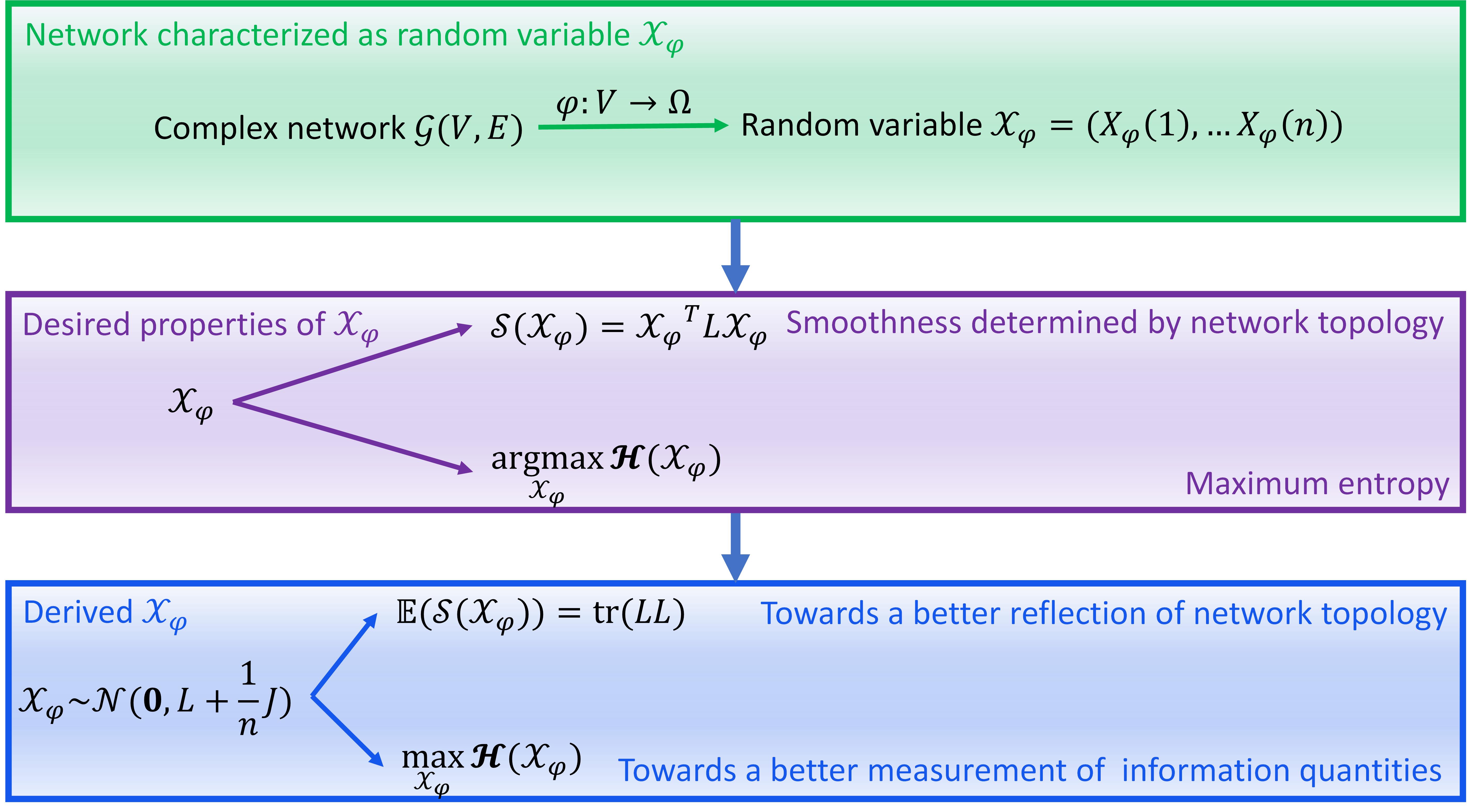}
\caption{\label{G1} Ideas of solving questions (I-II). We define a variable $\mathcal{X}_{\phi}$ on network $\mathcal{G}$ based on mapping $\phi$. An ideal definition of $\mathcal{X}_{\phi}$ is expected to make the smoothness of $\phi$ be completely determined by network topology. Meanwhile, it is expected to ensure the maximum entropy property of variable $\mathcal{X}_{\phi}$.} 
 \end{figure}

To properly measure information quantities (e.g., mutual information) between complex networks applying $\mathcal{X}_{\phi}$, the upper bounds of these quantities in $\mathcal{X}_{\phi}$ should not be too small. Otherwise, these quantities may be easily covered by noises in empirical data due to their small orders of magnitude. In the present study, we primarily focus on the Shannon entropy because extensive information upper bounds are related to it \cite{cover1999elements}. This idea inspires us to consider maximum entropy distribution problem \cite{cover1999elements} while defining mapping $\phi$ (see Fig. 1).

In sum, one way for solving questions (I-II) is to consider both the smoothness and maximum entropy properties of mapping $\phi$. Below, we suggest a potential solution.

\section{Gaussian variable defined by the function of Laplacian}\label{Sec3}
To avoid that smoothness $\mathcal{S}\left(\phi\right)$ in Eq. (\ref{EQ1}) diverges, random variable $\mathcal{X}_{\phi}$ is expected to have finite $1$-st and $2$-nd moments on each dimension. Please note that this setting has no explicit relation with the divergent $2$-nd moment of the degree distribution of a scale-free network \cite{albert2002statistical}. The finite moments of $\mathcal{X}_{\phi}$ are only proposed for ensuring the mathematical simplicity of $\mathcal{S}\left(\phi\right)$ (i.e., a divergent $\mathcal{S}\left(\phi\right)$ is meaningless in application). 

In our theory, we suggest a possible scheme
\begin{align}
     \mathbb{E}\left(\mathcal{X}_{\phi}\right)&=\mathbf{0},\label{EQ5}\\
     \mathbb{D}\left(\mathcal{X}_{\phi}\right)&=\Sigma\in\mathbb{R}^{n\times n},\label{EQ6}
\end{align}
where $\mathbf{0}=\left(0,\ldots,0\right)$ is a vector of zeros, $\mathbb{E}\left(\cdot\right)$ denotes the $1$-st moment, and $\mathbb{D}\left(\cdot\right)$ denotes the $2$-nd moment. Given Eqs. (\ref{EQ5}-\ref{EQ6}), we can reformulate Eq. (\ref{EQ1}) as 
\begin{align}
     \mathbb{E}\left(\mathcal{S}\left(\phi\right)\right)&=\sum_{\left(v_{i},v_{j}\right)\in E}\sqrt{W_{i,j}}\left(\Sigma_{ii}+\Sigma_{jj}-2\Sigma_{ij}\right),\label{EQ7}
\end{align}
where $W$, the weighted adjacent matrix, is predetermined by a given network. In Eq. (\ref{EQ7}), matrix $\Sigma$ is the only one adjustable term. To enable $\mathcal{X}_{\phi}$ to reflect the topology of $\mathcal{G}$, we suggest to choose matrix $\Sigma$ as
\begin{align}
     \Sigma=L+\frac{1}{n}J,\label{EQ8}
\end{align}
where $J$ is an all-one matrix. The motivation of the above definition lies in four aspects. First, although $L$ is a singular matrix, previous studies have proven that $L+\frac{1}{n}J$ is invertible if network $\mathcal{G}$ is connected \cite{xiao2003resistance,chebotarev2006proximity}
\begin{align}
     \left(L+\frac{1}{n}J\right)^{-1}=L^{\dagger}+\frac{1}{n}J,\label{EQ9}
\end{align}
where $\dagger$ denotes the Moore–Penrose pseudoinverse that satisfy $LL^{\dagger}L=L$ \cite{barata2012moore}. This property ensures the possibility to calculate numerous quantities defined with $ \Sigma^{-1}$ in Sec. \ref{Sec4}. Second, Eqs. (\ref{EQ8}-\ref{EQ9}) relate $\Sigma$ with $L^{\dagger}$ directly. The pseudoinverse Laplacian $L^{\dagger}$ is the \emph{reproducing kernel} of $\mathbb{H}\left(\mathcal{G}\right)$, the Hilbert space of real-valued functions over the node set $f_{\mathcal{G}}:V\rightarrow \mathbb{R}^{n}$ whose inner product is $\langle f_{\mathcal{G}},g_{\mathcal{G}}\rangle={f_{\mathcal{G}}}^T L g_{\mathcal{G}}$ \cite{herbster2005online,wahba1990spline}. Because $L^{\dagger}$ is unique for $\mathbb{H}\left(\mathcal{G}\right)$, we can confirm a unique $\mathbb{H}\left(\mathcal{G}\right)$ given $L^{\dagger}$. This property lays foundations for kernel tricks \cite{zhang2009reproducing,fukumizu2004dimensionality,zhou2003capacity} on network $\mathcal{G}$ when future studies explore machine learning tasks on random variable $\mathcal{X}_{\phi}$ (e.g., see kernel tricks in causality analysis \cite{chen2014causal,brodu2022discovering}). Third, Laplacian $L$ and its pseudoinverse $L^{\dagger}$ directly determine various topology properties of $\mathcal{G}$ (e.g., network coherence \cite{summers2015topology}, node importance \cite{van2017pseudoinverse}, and the number of spanning trees \cite{chung1997spectral}). Therefore, Eqs. (\ref{EQ8}-\ref{EQ9}) ensure the expressive ability of $\mathcal{X}_{\phi}$ about network topology. Fourth, we can apply Eq. (\ref{EQ4}) to derive the quadratic form
 \begin{align}
     \mathbb{E}\left(\mathcal{S}\left(\phi\right)\right)&=\mathbb{E}\left(\mathcal{X}_{\phi}\right)^{T}L\mathbb{E}\left(\mathcal{X}_{\phi}\right)+\operatorname{tr}\left(L\Sigma\right),\label{EQ10}\\
     &=\operatorname{tr}\left[L\left(L+\frac{1}{n}J\right)\right],\label{EQ11}
\end{align}
if Eq. (\ref{EQ8}) holds. Once network $\mathcal{G}$ is connected (i.e., there is only one zero eigenvalue in $\{ \lambda_{1},\ldots,\lambda_{n}\}$), we can further apply $LL^{\dagger}=L^{\dagger}L=I-\frac{1}{n}J$ \cite{gutman2004generalized,chebotarev2006proximity,van2017pseudoinverse}, where $I$ is the unit matrix, to derive
 \begin{align}
     \mathbb{E}\left(\mathcal{S}\left(\phi\right)\right)&=\operatorname{tr}\left[LL+L\left(I-L^{\dagger}L\right)\right],\label{EQ12}\\
     &=\operatorname{tr}\left(LL+L-LL^{\dagger}L\right),\label{EQ13}\\
     &=\operatorname{tr}\left(LL\right).\label{EQ14}
\end{align}
Eq. (\ref{EQ11}) and Eq. (\ref{EQ14}) suggest a benefit of Eq. (\ref{EQ8}) that we can control the expected smoothness of $\phi$ on a network completely by Laplacian $L$ (see Fig. 1).
  
To ensure the maximum entropy property, we need to analyze the maximum entropy distribution problem. Considering a random variable $\mathcal{X}_{\phi}\in\mathbb{R}^{n}$ with finite $1$-st and $2$-nd moments defined in Eqs. (\ref{EQ5}-\ref{EQ6}), we know 
 \begin{align}
     \mathcal{H}\left(\mathcal{X}_{\phi}\right)\leq\mathcal{H}\left(\mathcal{N}\left(\mathbf{0},\Sigma\right)\right),\label{EQ15}
\end{align}
where $\mathcal{H}\left(\cdot\right)$ denotes the Shannon entropy \cite{cover1999elements} and $\mathcal{N}\left(\mathbf{0},\Sigma\right)$ is the $n$-dimensional Gaussian distribution. Eq. (\ref{EQ15}) is derived from the fact that the maximum entropy distribution defined on $\mathbb{R}^{n}$ with given $1$-st and $2$-nd moments in Eqs. (\ref{EQ5}-\ref{EQ6}) is the Gaussian distribution \cite{cover1999elements}. Therefore, we define random variable $\mathcal{X}_{\phi}$ as
\begin{align}
     \mathcal{X}_{\phi}\sim\mathcal{N}\left(\mathbf{0},L+\frac{1}{n}J\right)\label{EQ16}
\end{align}
to reflect the topology of network $\mathcal{G}$ (see Fig. 1). In practice, we can readily derive the accurate entropy value
\begin{align}
    \mathcal{H}\left(\mathcal{X}_{\phi}\right)=\frac{n}{2}+\frac{n}{2}\ln\left(2\pi \right)+\frac{1}{2}\ln\left[\operatorname{det}\left(L+\frac{1}{n}J\right)\right],\label{EQ17}
\end{align}
where $\operatorname{det}\left(\cdot\right)$ denotes the determinant.

In sum, a possible solution of questions (I-II) is to represent $\mathcal{G}$ by a Gaussian variable in Eq. (\ref{EQ13}), which ensures the topology-dependent smoothness and maximum entropy properties of mapping $\phi$ (see Fig. 1). Such a variable is characterized by a function of the Laplacian operator \cite{biyikoglu2007laplacian}, whose precision matrix is $L^{\dagger}+\frac{1}{n}J$. 

Interestingly, we notice that Eq. (\ref{EQ8}) is similar with the graph signal characterization \cite{dong2016learning,kalofolias2016learn} derived by factor analysis \cite{bartholomew2011latent} and low-rank models \cite{tipping1999probabilistic,roweis1997algorithms}, which states that a Gaussian Markov random field representation (a special type of Gaussian variable) improves graph learning in practice \cite{dong2016learning,kalofolias2016learn}. The difference between Refs. \cite{dong2016learning,kalofolias2016learn} and our work lies in that they assume the covariance matrix as $\Sigma=L^{\dagger}$ while we define $\Sigma=L+\frac{1}{n}J$. This similarity suggests the validity of our ideas from the perspective of computation practice. In general, our definition can be treated as a variant of existing approaches \cite{dong2016learning,kalofolias2016learn}. Our main progress lies in that we offer a theoretical explanation for the mechanisms underlying the successes of these engineering practices from the perspectives of topology-dependent smoothness and maximum entropy. In Sec. \ref{Sec8-2}, we present a comprehensive comparison between our work and previous studies \cite{dong2016learning,kalofolias2016learn}.

\section{Analytic metrics of network relations}\label{Sec4}
After representing two networks $\mathcal{G}_{a}$ and $\mathcal{G}_{b}$ by variables $\mathcal{X}_{\phi}^{a}=\left(X_{\phi}^{a}\left(1\right),\ldots,X_{\phi}^{a}\left(n\right)\right)\sim\mathcal{N}\left(\mathbf{0},L\left(a\right)+\frac{1}{n}J\right)$ and $\mathcal{X}_{\phi}^{b}=\left(X_{\phi}^{b}\left(1\right),\ldots,X_{\phi}^{b}\left(n\right)\right)\sim\mathcal{N}\left(\mathbf{0},L\left(b\right)+\frac{1}{n}J\right)$, we can develop analytic metrics of network relations. 

 \begin{figure}[t!]
\includegraphics[width=1\columnwidth]{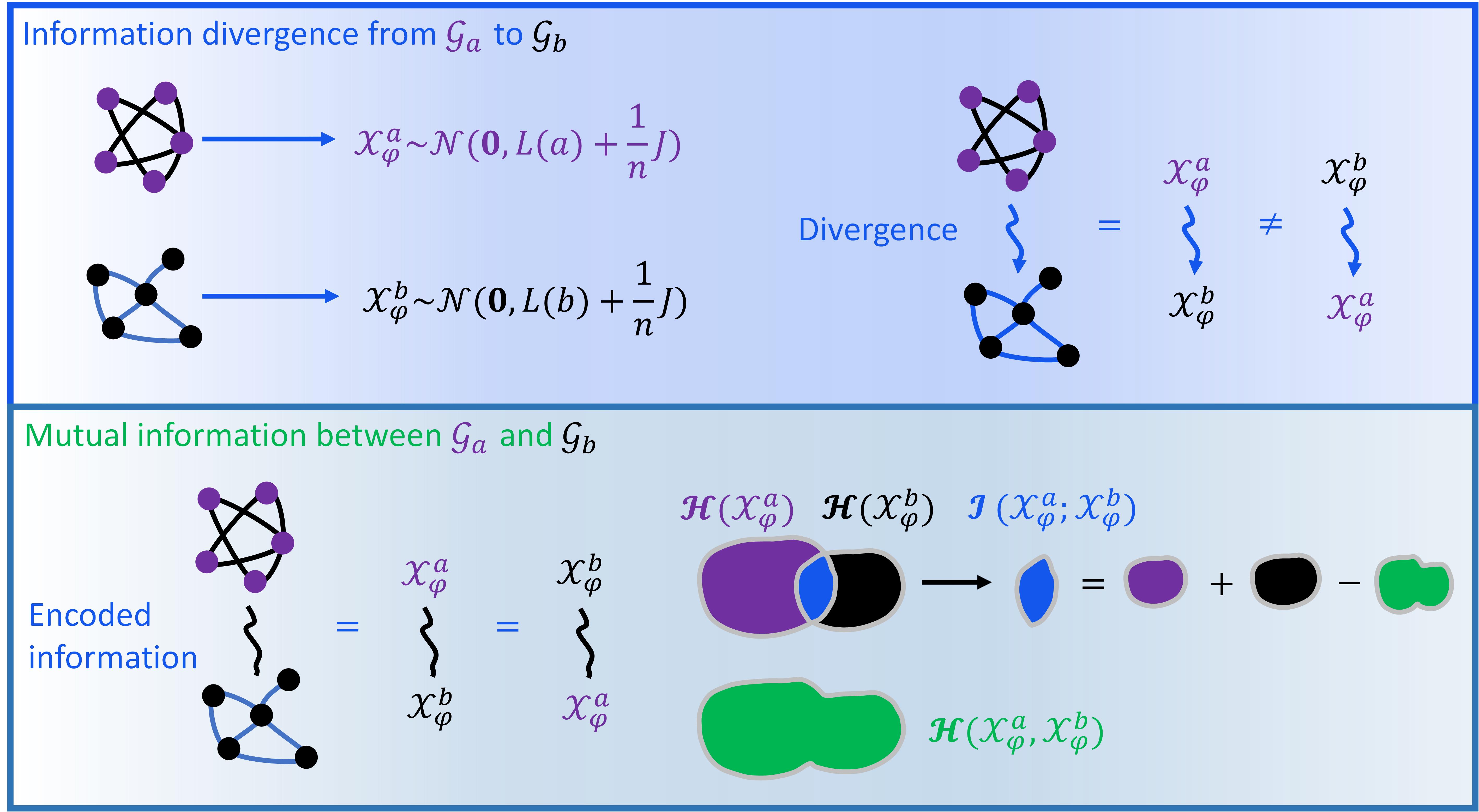}
\caption{\label{G2} Information divergence and mutual information between complex networks. The information divergence from network $\mathcal{G}_{a}$ to network $\mathcal{G}_{b}$, defined by $ \mathcal{D}\left(\mathcal{X}_{\phi}^{a}\Vert\mathcal{X}_{\phi}^{b}\right)$ in Eq. (\ref{EQ20}), is not necessarily equal to the information divergence from network $\mathcal{G}_{b}$ to network $\mathcal{G}_{a}$ (upper parallel). The mutual information between networks $\mathcal{G}_{a}$ and $\mathcal{G}_{b}$, defined by $\mathcal{I}\left(\mathcal{X}_{\phi}^{a};\mathcal{X}_{\phi}^{b}\right)$ in Eq. (\ref{EQ21}), can be understood as the shared part of entropy quantities $\mathcal{H}\left(\mathcal{X}_{\phi}^{a}\right)$ and $\mathcal{H}\left(\mathcal{X}_{\phi}^{b}\right)$ (bottom parallel).} 
 \end{figure}

\subsection{Encoding: information divergence and mutual information}\label{Sec4-1}
For information divergence (or referred to as the Kullback–Leibler divergence \cite{cover1999elements}), we can formulate it in a conventional form
\begin{align}
   \mathcal{D}\left(\mathcal{X}_{\phi}^{a}\Vert\mathcal{X}_{\phi}^{b}\right)&=\mathbb{E}_{\rho_{a}}\left[\log\left(\rho_{a}\right)-\log\left(\rho_{b}\right)\right],\label{EQ18}
\end{align}
where $\rho_{a}$ and $\rho_{b}$ are probability densities of $\mathcal{X}_{\phi}^{a}$ and $\mathcal{X}_{\phi}^{b}$, respectively (see Fig. 2). Because $\mathcal{X}_{\phi}^{a}$ and $\mathcal{X}_{\phi}^{b}$ are Gaussian variables, we can derive
\begin{align}
   \log\left(\rho_{a}\right)=&-\frac{1}{2}\log\left[\left(2\pi\right)^{n}\operatorname{det}\left(L\left(a\right)+\frac{1}{n}J\right)\right]\notag\\&-\frac{1}{2}\left[\mathcal{X}_{\phi}^{a}\right]^{T}\left(L^{\dagger}\left(a\right)+\frac{1}{n}J\right)\mathcal{X}_{\phi}^{a},\label{EQ19}
\end{align}
which readily leads to
\begin{align}
   \mathcal{D}\left(\mathcal{X}_{\phi}^{a}\Vert\mathcal{X}_{\phi}^{b}\right)=&\frac{1}{2}\Bigg[\operatorname{tr}\left[\left(L^{\dagger}\left(b\right)+\frac{1}{n}J\right)\left(L\left(a\right)+\frac{1}{n}J\right)\right]\notag\\&-n+\ln\frac{\operatorname{det}\left(L\left(b\right)+\frac{1}{n}J\right)}{\operatorname{det}\left(L\left(a\right)+\frac{1}{n}J\right)}\Bigg].\label{EQ20}
\end{align}
Eq. (\ref{EQ20}) measures the directional difference between the topology of $\mathcal{G}_{a}$ and $\mathcal{G}_{b}$. The difference is directional since $\mathcal{D}\left(\mathcal{X}_{\phi}^{a}\Vert\mathcal{X}_{\phi}^{b}\right)\neq\mathcal{D}\left(\mathcal{X}_{\phi}^{b}\Vert\mathcal{X}_{\phi}^{a}\right)$ (see Fig. 2). An important property of the information divergence defined in Eq. (\ref{EQ20}) lies in that it is completely determined by the Laplacian spectra of two networks. Therefore, it is less suitable for comparing between iso-spectral networks (i.e., networks can share a same Laplacian spectrum but have different network topology properties).

For mutual information $\mathcal{I}\left(\mathcal{X}_{\phi}^{a};\mathcal{X}_{\phi}^{b}\right)$ that quantifies the topology information of network $\mathcal{G}_{a}$ encoded by network $\mathcal{G}_{b}$, we can calculate (see Fig. 2)
\begin{align}
   \mathcal{I}\left(\mathcal{X}_{\phi}^{a};\mathcal{X}_{\phi}^{b}\right)&=\mathcal{H}\left(\mathcal{X}_{\phi}^{a}\right)+\mathcal{H}\left(\mathcal{X}_{\phi}^{b}\right)-\mathcal{H}\left(\mathcal{X}_{\phi}^{a},\mathcal{X}_{\phi}^{b}\right),\label{EQ21}
\end{align}
where $\mathcal{H}\left(\mathcal{X}_{\phi}^{a}\right)$ and $\mathcal{H}\left(\mathcal{X}_{\phi}^{b}\right)$ can be measured based on Eq. (\ref{EQ17}). A challenge in Eq. (\ref{EQ21}) lies in that $\mathcal{H}\left(\mathcal{X}_{\phi}^{a},\mathcal{X}_{\phi}^{b}\right)$ is non-trivial for analytic derivations unless variables $X_{\phi}^{a}$ and $X_{\phi}^{b}$ are jointly Gaussian (this enables $\mathcal{H}\left(\mathcal{X}_{\phi}^{a},\mathcal{X}_{\phi}^{b}\right)$ to be defined by Eq. (\ref{EQ17}) as well). In more general cases where we do not know whether $X_{\phi}^{a}$ and $X_{\phi}^{b}$ are jointly Gaussian or not, we generate samples of $\mathcal{X}_{\phi}^{a}\sim\mathcal{N}\left(\mathbf{0},L\left(a\right)+\frac{1}{n}J\right)$ and $\mathcal{X}_{\phi}^{b}\sim\mathcal{N}\left(\mathbf{0},L\left(b\right)+\frac{1}{n}J\right)$ by inverse transform sampling \cite{olver2013fast} to estimate $\mathcal{H}\left(\mathcal{X}_{\phi}^{a},\mathcal{X}_{\phi}^{b}\right)$ using the Kozachenko-Leonenko estimator of Shannon entropy \cite{kozachenko1987sample,kraskov2004estimating}. This approach enables us to derive mutual information in real situations.       

\begin{figure}[t!]
\includegraphics[width=1\columnwidth]{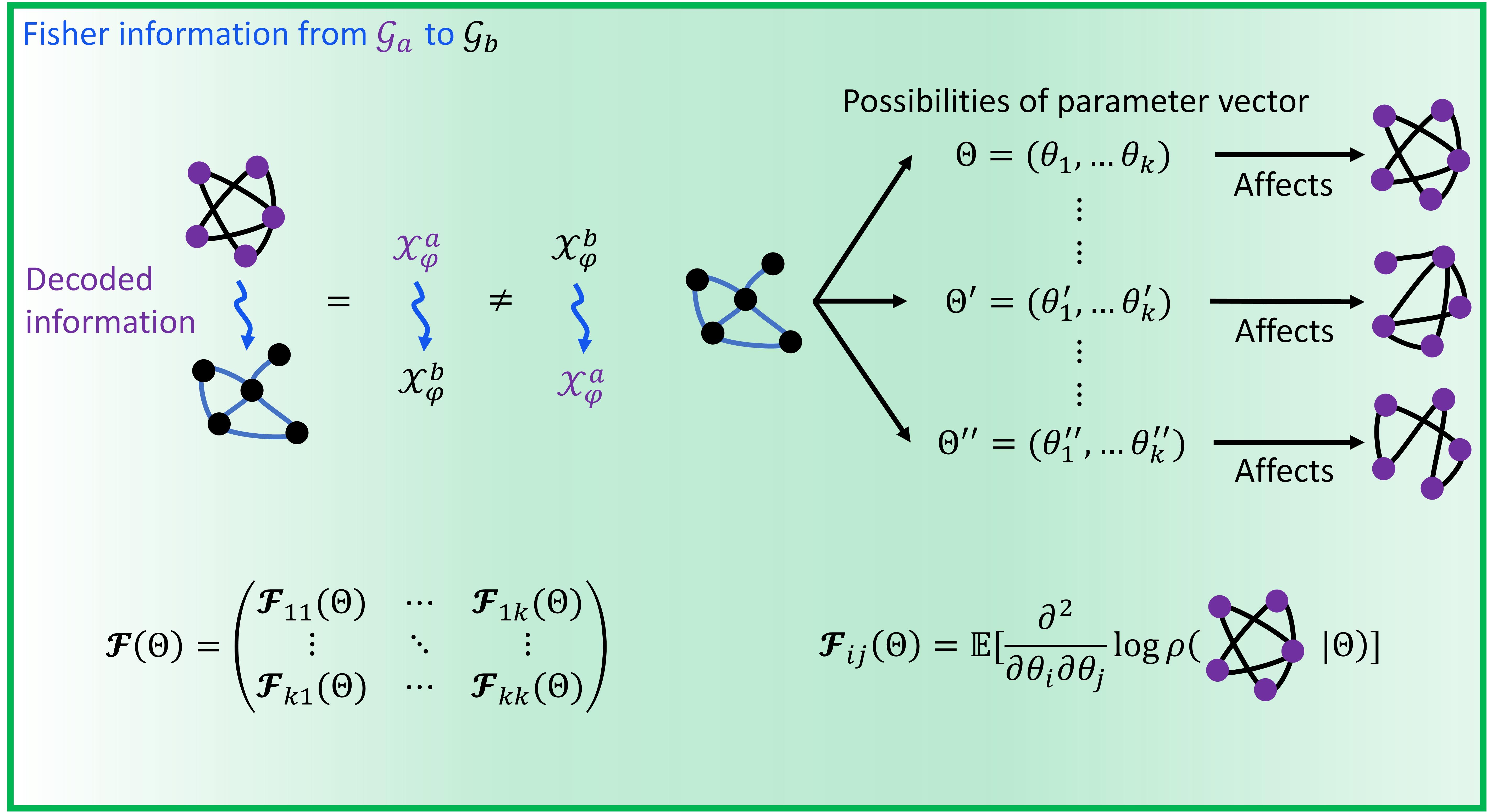}
\caption{\label{G3} Fisher information between complex networks. The $\left(i,j\right)$ element in the Fisher information matrix, denoted by $\mathcal{F}_{ij}\left(\Theta\right)$ in Eq. (\ref{EQ22}), can be understood as the information that network $\mathcal{G}_{a}$ carries about the parameter vector $\Theta$ controlled by network $\mathcal{G}_{b}$. The information is contained in $\rho_{a}\left(\cdot\mid\Theta\right)$, the probability distribution of variable $\mathcal{X}_{\phi}^{a}$ that is affected by $\Theta$. The information quantity can be understood as the mean sensitivity of $\rho_{a}\left(\cdot\mid\Theta\right)$ towards the variation of $\Theta$.}
 \end{figure}

\subsection{Decoding: Fisher information}\label{Sec4-2}
For Fisher information, we assume that a parameter vector $\Theta=\left(\theta_{1},\ldots,\theta_{k}\right)\in\mathbb{R}^{k}$ controlled by network $\mathcal{G}_{b}$ can affect the Laplacian $L\left(a\right)$ of network $\mathcal{G}_{a}$. Fisher information measures how precisely we can decode the topology information of $\mathcal{G}_{a}$ from $\mathcal{G}_{b}$ according to parameter vector $\Theta$. We denote $\mathcal{X}_{\phi}^{a}\sim\mathcal{N}\left(\mathbf{0},L^{\dagger}\left(a\mid\Theta\right)\right)$ as the Gaussian variable given parameter vector $\Theta$ (see Fig. 3). Then we can have a special form of Fisher information matrix depending on the covariance matrix \cite{mardia1984maximum,malago2015information} 
\begin{align}
   \mathcal{F}_{ij}\left(\Theta\right)=&-\int \rho_{a}\left(\chi;\Theta\right)\left\{\frac{\partial^{2}}{\partial \theta_{i}\theta_{j}}\log\left[\rho_{a}\left(\chi\mid\Theta^{\prime}\right)\right]\Bigg\vert_{\Theta^{\prime}=\Theta}\right\}\mathsf{d}\chi,\label{EQ22}\\
   =&\frac{1}{2}\operatorname{tr}\Bigg[\left(L^{\dagger}\left(a\mid\Theta\right)+\frac{1}{n}J\right)\frac{\partial L\left(a\mid\Theta\right)}{\partial\theta_{i}}\notag\\&\times\left(L^{\dagger}\left(a\mid\Theta\right)+\frac{1}{n}J\right)\frac{\partial L\left(a\mid\Theta\right)}{\partial\theta_{j}}\Bigg]\label{EQ23},
\end{align}
where $\rho_{a}\left(\cdot\mid\Theta\right)$ is the probability density of $\mathcal{X}_{\phi}^{a}$ given $\Theta$ (see Fig. 3). We define
\begin{align}
   \dfrac{\partial L\left(a;\Theta\right)}{\partial\theta_{i}}&=
   \left[\begin{array}{ccc}
   \dfrac{\partial L_{11}\left(a\mid\Theta\right)}{\partial\theta_{i}} & \ldots & \dfrac{\partial L_{1n}\left(a\mid\Theta\right)}{\partial\theta_{i}} \\ \vdots & \ddots & \vdots\\
   \dfrac{\partial L_{n1}\left(a\mid\Theta\right)}{\partial\theta_{i}} & \ldots & \dfrac{\partial L_{nn}\left(a\mid\Theta\right)}{\partial\theta_{i}}
   \end{array}\right]. \label{EQ24}
\end{align}
The expectation vector does not occur in Eq. (\ref{EQ23}) since $\mathcal{X}_{\phi}^{a}$ has zero expectation on each dimension. In application, one can further calculate Fisher information quantity, $\operatorname{tr}\left[\mathcal{F}\left(\Theta\right)\right]$, as a metric of decoding precision.

\subsection{Causality: Granger causality and conditional mutual information}\label{Sec4-3}

To this point, we have analytically derived information divergence, mutual information, and Fisher information between complex networks. These metrics lay the foundation of encoding and decoding analyses on network ensembles. Compared with these analyses, causality is more technically non-trivial to study between networks because it is previously limited to time series \cite{hlavavckova2007causality}. Although dynamic networks feature time domain evolution \cite{casteigts2012time,zimmermann2004coevolution,hill2010dynamic}, most networks lack a well-defined concept of time (e.g., networks may be static). To develop applicable causality metrics for arbitrary networks, we explore possible generalization of the mainstream causality metrics, such as transfer entropy \cite{schreiber2000measuring,staniek2008symbolic,tian2021fourier} and Granger causality \cite{shojaie2022granger,friston2014granger,bueso2020explicit,marinazzo2008kernel,marinazzo2008kernel}, from time domain to graph domain. Please note that transfer entropy is equivalent to conditional mutual information \cite{cover1999elements} if we do not apply the terminology of time series. For the sake of clarity, we only use conditional mutual information as its name in our framework.

Let us begin with a classic form of Granger causality analyzed by regression models. Our basic idea is to consider a random partition on network $\mathcal{G}_{a}$ to divide it into two sub-networks, $\mathcal{G}_{a}^{\circ}$ and $\mathcal{G}_{a}^{\star}$. This is equivalent to dividing random variable $\mathcal{X}_{\phi}^{a}=\left(X_{\phi}^{a}\left(1\right),\ldots,X_{\phi}^{a}\left(n\right)\right)$ into two sub-vectors of multivariate Gaussian random variables $\mathcal{X}_{\phi}^{a,\circ}$ and $\mathcal{X}_{\phi}^{a,\star}$. Without loss of generality, we set $\mathcal{X}_{\phi}^{a,\circ}=\left(X_{\phi}^{a}\left(1\right),\ldots,X_{\phi}^{a}\left(k\right)\right)\sim\mathcal{N}\left(\mathbf{0},L\left(a,\circ\right)+\frac{1}{n}J\right)$ and $\mathcal{X}_{\phi}^{a,\star}=\left(X_{\phi}^{a}\left(k+1\right),\ldots,X_{\phi}^{a}\left(n\right)\right)\sim\mathcal{N}\left(\mathbf{0},L\left(a,\star\right)+\frac{1}{n}J\right)$, where we define
\begin{align}
  L\left(a,\circ\right)&=
   \left[\begin{array}{ccc}
   L_{11}\left(a\right) & \ldots & L_{1k}\left(a\right) \\ \vdots & \ddots & \vdots\\
   L_{k1}\left(a\right) & \ldots & L_{kk}\left(a\right)
   \end{array}\right],\label{EQ25}\\
   L\left(a,\star\right)&=
   \left[\begin{array}{ccc}
   L_{\left(k+1\right)\left(k+1\right)}\left(a\right) & \ldots & L_{\left(k+1\right)n}\left(a\right) \\ \vdots & \ddots & \vdots\\
   L_{n\left(k+1\right)}\left(a\right) & \ldots & L_{nn}\left(a\right)
   \end{array}\right],\label{EQ26}
\end{align}
and we can represent $L\left(a\right)$ in a block matrix form $L\left(a\right)=\left[\begin{array}{cc}
   L\left(a,\circ\right)  & L\left(a,\bigtriangleup\right) \\ 
   L\left(a,\bigtriangledown\right) & L\left(a,\star\right)
   \end{array}\right]$. Then, we use $\mathcal{X}_{\phi}^{a,\circ}$ to predict $\mathcal{X}_{\phi}^{a,\star}$ with a linear model
\begin{align}
   \mathcal{X}_{\phi}^{a,\star}&=\beta+\mathcal{X}_{\phi}^{a,\circ}A+\varepsilon,\label{EQ27}
\end{align}
where $A$ denotes the regression coefficient matrix, $\beta$ is a constant vector, and $\varepsilon$ measures regression residuals. Meanwhile, we can also use $\mathcal{X}_{\phi}^{a,\circ}$ and $\mathcal{X}_{\phi}^{b}$ to predict $\mathcal{X}_{\phi}^{a,\star}$
\begin{align}
   \mathcal{X}_{\phi}^{a,\star}&=\beta^{\prime}+\left(\mathcal{X}_{\phi}^{a,\circ}\oplus\mathcal{X}_{\phi}^{b}\right)A^{\prime}+\varepsilon^{\prime},\label{EQ28}
\end{align}
where we have applied notion $\oplus$ to denote the concatenation of two vectors, i.e., $\mathcal{X}_{\phi}^{a,\circ}\oplus\mathcal{X}_{\phi}^{b}=\left(X_{\phi}^{a}\left(1\right),\ldots,X_{\phi}^{a}\left(k\right),X_{\phi}^{b}\left(1\right),\ldots,X_{\phi}^{b}\left(n\right)\right)$ (see Fig. 4). According to Refs. \cite{barnett2009granger,davidson2004econometric}, the ordinary least squares regression for Eqs. (\ref{EQ27}-\ref{EQ28}) is suggested to minimize the determinant of covariance matrix of residuals (referred to as the generalized variance). The covariance matrices of residuals for Eqs. (\ref{EQ27}-\ref{EQ28}) are 
\begin{align}
   \Sigma\left(\varepsilon\right)&=\Sigma\left(\mathcal{X}_{\phi}^{a,\star}\big\vert\mathcal{X}_{\phi}^{a,\circ}\right),\label{EQ29}\\
   \Sigma\left(\varepsilon^{\prime}\right)&=\Sigma\left(\mathcal{X}_{\phi}^{a,\star}\big\vert\mathcal{X}_{\phi}^{a,\circ}\oplus\mathcal{X}_{\phi}^{b}\right).\label{EQ30}
\end{align}
The Granger causality can be defined as the natural logarithmic ratio between the determinant values of Eqs. (\ref{EQ29}-\ref{EQ30}) \cite{barnett2009granger}
\begin{align}
   \mathcal{T}_{G}\left(\mathcal{X}_{\phi}^{b}\rightarrow\mathcal{X}_{\phi}^{a}\right)&=\Bigg\langle\ln\left[\frac{\operatorname{det}\left(\Sigma\left(\varepsilon\right)\right)}{\operatorname{det}\left(\Sigma\left(\varepsilon^{\prime}\right)\right)}\right]\Bigg\rangle,\label{EQ31}
\end{align}
where the average $\langle\cdot\rangle$ is implemented across multiple randomly generated configurations of $\mathcal{X}_{\phi}^{a,\star}$ and $\mathcal{X}_{\phi}^{a,\circ}$ (i.e., we can randomly select $h$ configurations of $\mathcal{X}_{\phi}^{a,\star}$ and $\mathcal{X}_{\phi}^{a,\circ}$ to calculate $h$ values of $\ln\left[\frac{\Sigma\left(\varepsilon\right)}{\Sigma\left(\varepsilon^{\prime}\right)}\right]$ and average across them to derive Eq. (\ref{EQ31})). Please see Fig. 4 for illustrations. Because $\mathcal{X}_{\phi}^{a,\star}$ and $\mathcal{X}_{\phi}^{a,\circ}$ are jointly Gaussian \cite{davidson2004econometric,barnett2009granger}, we can readily derive
\begin{align}
   \Sigma\left(\varepsilon\right)=&L\left(a,\circ\right)+\frac{1}{n}J-\left(L\left(a,\bigtriangleup\right)+\frac{1}{n}J\right)\notag\\&\times\left(L\left(a,\star\right)+\frac{1}{n}J\right)^{-1}\left(L\left(a,\bigtriangledown\right)+\frac{1}{n}J\right).\label{EQ32}
\end{align}
However, $\Sigma\left(\varepsilon^{\prime}\right)$ can not be analytically derived by the Laplacians of networks unless we relax conditions (e.g., let $\mathcal{X}_{\phi}^{a,\star}$, $\mathcal{X}_{\phi}^{a,\circ}$, and $\mathcal{X}_{\phi}^{b}$ be jointly Gaussian as well). Similar to the situation in Eq. (\ref{EQ21}), we suggest that one can generate samples of $\mathcal{X}_{\phi}^{a,\star}$, $\mathcal{X}_{\phi}^{a,\circ}$, and $\mathcal{X}_{\phi}^{b}$ by inverse transform sampling \cite{olver2013fast} to estimate $\Sigma\left(\varepsilon^{\prime}\right)$ in practice. Please note that our derivations presented above do not require any knowledge about node alignment (i.e., node $v_{i}$ in network $\mathcal{G}_{a}$ corresponds to node $v_{j}$ in network $\mathcal{G}_{b}$). 

 \begin{figure}[t!]
\includegraphics[width=1\columnwidth]{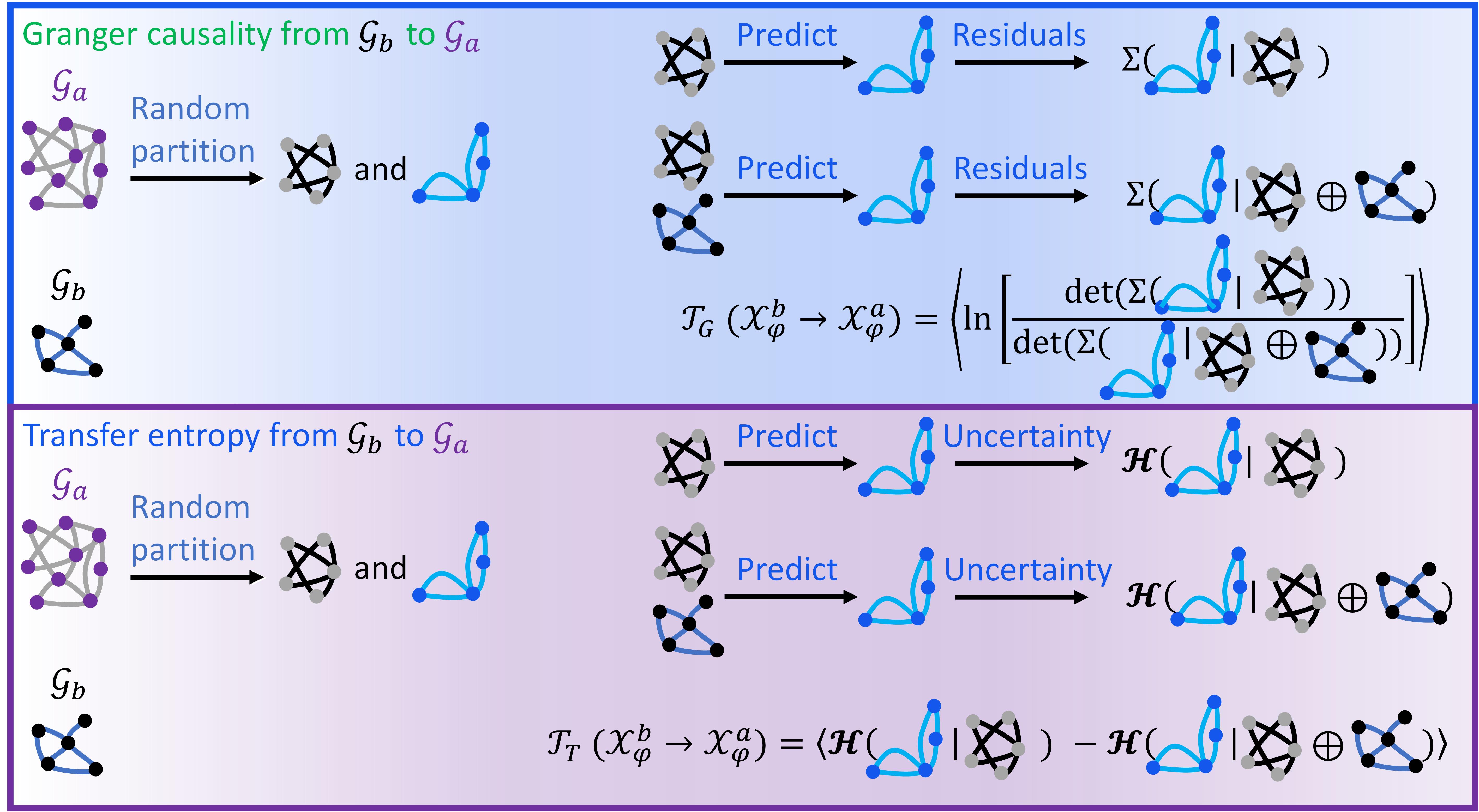}
\caption{\label{G4} Granger causality and conditional mutual information between complex networks. To quantify the causal effects from $\mathcal{G}_{b}$ to $\mathcal{G}_{a}$, we first do random partition on network $\mathcal{G}_{a}$ to obtain two sub-networks. Then we predict the topology properties of one sub-network based on another sub-network. The prediction may involve with residuals (in terms of Granger causality) or uncertainty (in terms of conditional mutual information). The causal effects from $\mathcal{G}_{b}$ to $\mathcal{G}_{a}$ are reflected by the reduction of residuals and uncertainty after we include the information of $\mathcal{G}_{b}$ into the prediction. By repeating random partition and prediction, we measure Granger causality and conditional mutual information in terms of the average reduction of residuals and uncertainty.} 
 \end{figure}

Then, we turn to formulating conditional mutual information \cite{cover1999elements} (i.e., the counterpart of transfer entropy defined between networks)
\begin{align}
   &\mathcal{T}_{T}\left(\mathcal{X}_{\phi}^{b}\rightarrow\mathcal{X}_{\phi}^{a}\right)\notag\\=&\Big\langle\mathcal{I}\left(\mathcal{X}_{\phi}^{a,\star};\mathcal{X}_{\phi}^{b}\big\vert\mathcal{X}_{\phi}^{a,\circ}\right)\Big\rangle\label{EQ33}\\=&\Big\langle\mathcal{H}\left(\mathcal{X}_{\phi}^{a,\star}\big\vert\mathcal{X}_{\phi}^{a,\circ}\right)-\mathcal{H}\left(\mathcal{X}_{\phi}^{a,\star}\big\vert\mathcal{X}_{\phi}^{a,\circ}\oplus\mathcal{X}_{\phi}^{b}\right)\Big\rangle.\label{EQ34}
\end{align}
Please see Fig. 4 for illustrations. Similar to Eq. (\ref{EQ32}), we can derive
\begin{align}
   \mathcal{H}\left(\mathcal{X}_{\phi}^{a,\star}\big\vert\mathcal{X}_{\phi}^{a,\circ}\right)=&\mathcal{H}\left(\mathcal{X}_{\phi}^{a,\star}\oplus\mathcal{X}_{\phi}^{a,\circ}\right)-\mathcal{H}\left(\mathcal{X}_{\phi}^{a,\circ}\right),\label{EQ35}\\=&\frac{n-k}{2}+\frac{n-k}{2}\log\left(2\pi \right)\notag\\&+\frac{1}{2}\log\left[\frac{\operatorname{det}\left(L\left(a\right)+\frac{1}{n}J\right)}{\operatorname{det}\left(L\left(a,\circ\right)+\frac{1}{n}J\right)}\right]\label{EQ36}
\end{align}
because $\mathcal{X}_{\phi}^{a,\star}$ and $\mathcal{X}_{\phi}^{a,\circ}$ are jointly Gaussian. Similar to the cases in Eq. (\ref{EQ21}) and $\Sigma\left(\varepsilon^{\prime}\right)$, we can not derive a general expression of $\mathcal{H}\left(\mathcal{X}_{\phi}^{a,\star}\big\vert\mathcal{X}_{\phi}^{a,\circ}\oplus\mathcal{X}_{\phi}^{b}\right)=\mathcal{H}\left(\mathcal{X}_{\phi}^{a},\mathcal{X}_{\phi}^{b}\right)-\mathcal{H}\left(\mathcal{X}_{\phi}^{a,\circ}\oplus\mathcal{X}_{\phi}^{b}\right)$ directly. In practice, we can resolve this challenge by inverse transform sampling \cite{olver2013fast} the Kozachenko-Leonenko estimator of Shannon entropy \cite{kozachenko1987sample,kraskov2004estimating}. 

The causality metrics considered here, such as $\mathcal{T}_{G}\left(\mathcal{X}_{\phi}^{b}\rightarrow\mathcal{X}_{\phi}^{a}\right)$ and $\mathcal{T}_{T}\left(\mathcal{X}_{\phi}^{b}\rightarrow\mathcal{X}_{\phi}^{a}\right)$, should be referred to as apparent causality metrics according to Ref. \cite{lizier2010differentiating} (one can see similar apparent causality metrics for time series in Refs. \cite{schreiber2000measuring,staniek2008symbolic,tian2021fourier,barnett2012transfer,shojaie2022granger,friston2014granger,bueso2020explicit,marinazzo2008kernel,marinazzo2008kernel}). During predicting the topology properties of sub-network $\mathcal{G}_{a}^{\star}$ by the characteristics of sub-network $\mathcal{G}_{a}^{\circ}$, these apparent causality metrics mainly reflect how the residuals and uncertainty of prediction are reduced by including the information of network $\mathcal{G}_{b}$. A higher reduction degree means that $\mathcal{G}_{b}$ contains more information about $\mathcal{G}_{a}^{\star}$ on average, implying that network $\mathcal{G}_{b}$ is more similar to network $\mathcal{G}_{a}$. Therefore, these metrics can be principally used for network comparison. Compared with  the information divergence, mutual information, and Fisher information derived in Eqs. (\ref{EQ18}-\ref{EQ24}), these apparent causality metrics convey more knowledge about the information flow from $\mathcal{G}_{b}$ to $\mathcal{G}_{a}$.

To derive complete causality metrics that reflect causal relations more precisely (e.g., enable $\mathcal{T}_{G}$ and $\mathcal{T}_{T}$ approximate causal information flow \cite{ay2008information}), one need to consider $\mathcal{T}_{G}\left(\mathcal{X}_{\phi}^{b}\rightarrow\mathcal{X}_{\phi}^{a}\big\vert \mathcal{Y}\right)$ and $\mathcal{T}_{T}\left(\mathcal{X}_{\phi}^{b}\rightarrow\mathcal{X}_{\phi}^{a}\big\vert \mathcal{Y}\right)$ given a reference variable $\mathcal{Y}$. Because the details of introducing $\mathcal{Y}$ have been comprehensively explored in Refs. \cite{lizier2010differentiating,hlavackova2011equivalence,barnett2009granger,lizier2008local} and these details do not imply critical challenges for mathematical derivations, we no longer repeat their analyses here. One can combine Refs. \cite{lizier2010differentiating,hlavackova2011equivalence,barnett2009granger,lizier2008local} and our framework to derive $\mathcal{T}_{G}\left(\mathcal{X}_{\phi}^{b}\rightarrow\mathcal{X}_{\phi}^{a}\big\vert \mathcal{Y}\right)$ and $\mathcal{T}_{T}\left(\mathcal{X}_{\phi}^{b}\rightarrow\mathcal{X}_{\phi}^{a}\big\vert \mathcal{Y}\right)$ between networks.

\section{Generalization of analytic metrics}\label{Sec5}
One may notice that our derivations in Sec. \ref{Sec4} are shown in a case where $\mathcal{X}_{\phi}^{a}$ and $\mathcal{X}_{\phi}^{b}$ are both $n$-dimensional, meaning that networks $\mathcal{G}_{a}$ and $\mathcal{G}_{b}$ both contain $n$ nodes. This limitation arises from the fact that we need to calculate $\left(L^{\dagger}\left(b\right)+\frac{1}{n}J\right)\left(L\left(a\right)+\frac{1}{n}J\right)$ in information divergence. The definitions of mutual information, Fisher information, Granger causality, and conditional mutual information have no such a limitation and are generally applicable to arbitrary cases.

 \begin{figure}[b!]
\includegraphics[width=1\columnwidth]{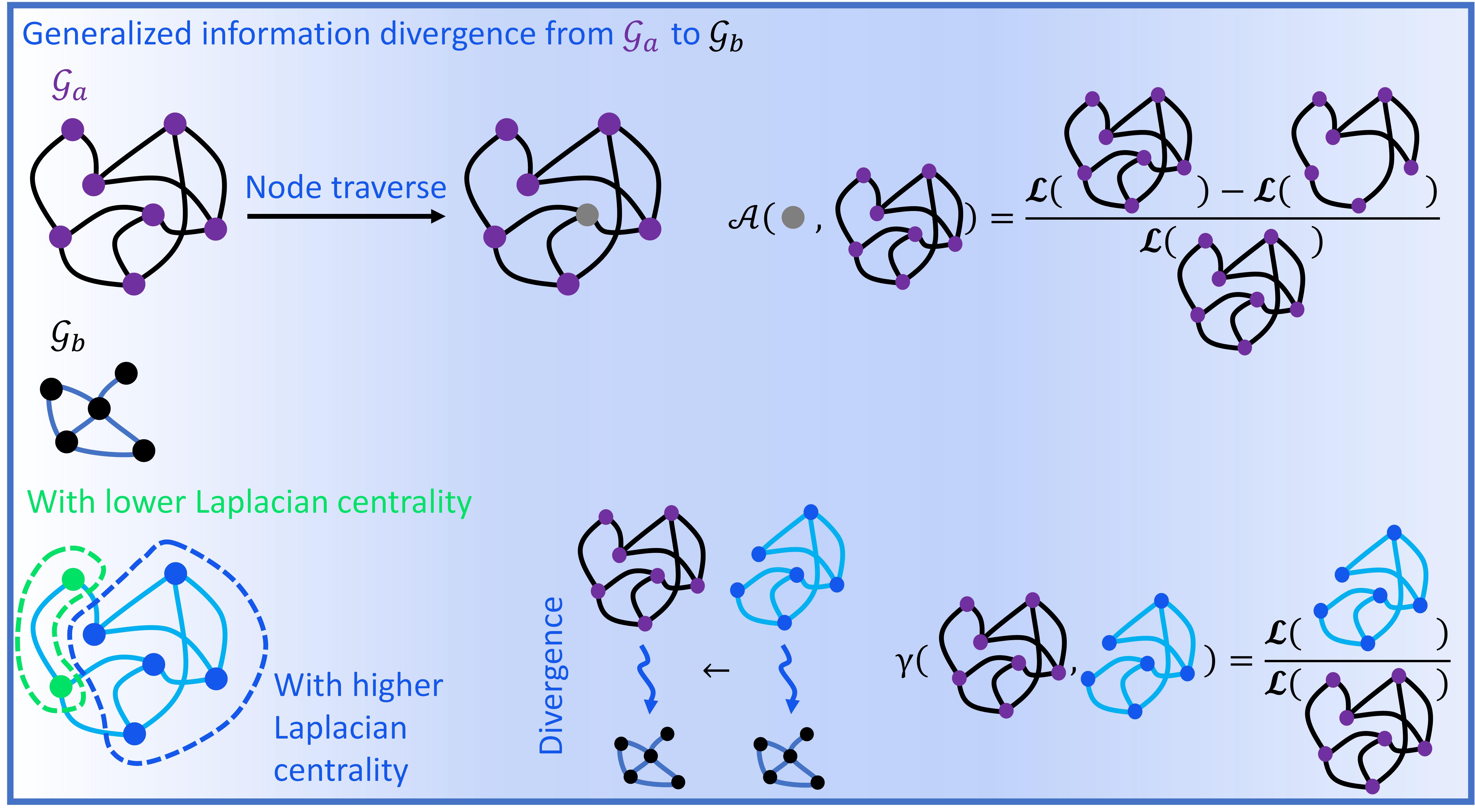}
\caption{\label{G5} The generalization of information divergence. When networks $\mathcal{G}_{a}$ and $\mathcal{G}_{b}$ have different sizes (e.g., network $\mathcal{G}_{a}$ contains more nodes), the information divergence defined in Eq. (\ref{EQ20}) can not be calculated directly. Consequently, we need to measure the importance of nodes in $\mathcal{G}_{a}$ from the perspective of network topology. In Eq. (\ref{EQ38}), we introduce $\mathcal{A}\left(\cdot,\mathcal{G}_{a}\right)$, the importance measurement based on Laplacian energy, as a possible approach. After assigning the importance of each node in $\mathcal{G}_{a}$, we can exclude the nodes with relatively lower importance to make the filtered $\mathcal{G}_{a}$ have the same size as $\mathcal{G}_{b}$. Information divergence from the filtered $\mathcal{G}_{a}$ to $\mathcal{G}_{b}$ can be measured and treated as an approximation of the information divergence from the original $\mathcal{G}_{a}$ to $\mathcal{G}_{b}$. The rationality $\gamma$ of this approximation can be measured as the fraction of the lost Laplacian energy in the original Laplacian energy.} 
 \end{figure}

In practice, we frequently need to analyze relations between complex networks with distinct sizes (number of nodes). To make our information divergence applicable to these networks, we suggest a practical solution based on Laplacian energy. Let us consider a case where $\mathcal{X}_{\phi}^{a}$ and $\mathcal{X}_{\phi}^{b}$ are $m$-dimensional and $n$-dimensional, respectively. Without loss of generality, we primarily discuss the case where $m>n$. The Laplacian energy of network $\mathcal{G}_{a}\left(V_{a},E_{a}\right)$ is defined as \cite{gutman2006laplacian,baruah2017comparative,yang2017parameter,qi2012laplacian}
\begin{align}
    \mathcal{L}\left(\mathcal{G}_{a}\right)&=\sum_{i=1}^{m}\lambda_{i}^{2}=\operatorname{tr}\left(L\left(a\right)L\left(a\right)\right),\label{EQ37}
\end{align}
where $\left(\lambda_{1},\ldots,\lambda_{m}\right)$ are the eigenvalues of $L\left(a\right)$. Note that Eq. (\ref{EQ37}) is equivalent to Eq. (\ref{EQ14}). Based on Eq. (\ref{EQ37}), we can measure the importance of each node $v_{i}$ in maintaining topology properties of $\mathcal{G}_{a}$ by Laplacian centrality \cite{baruah2017comparative,qi2012laplacian}
\begin{align}
    \mathcal{A}\left(v_{i},\mathcal{G}_{a}\right)&=\frac{\mathcal{L}\left(\mathcal{G}_{a}\right)-\mathcal{L}\left(\mathcal{G}_{a}/\{v_{i}\}\right)}{\mathcal{L}\left(\mathcal{G}_{a}\right)},\label{EQ38}
\end{align}
where $\mathcal{G}_{a}/\{v_{i}\}$ means deleting node $v_{i}$ from network $\mathcal{G}_{a}$. Note that we have $\mathcal{L}\left(\mathcal{G}_{a}\right)\geq\mathcal{L}\left(\mathcal{G}_{a}/\{v_{i}\}\right)$, where the equality holds if and only if $v_{i}$ is an isolate node (i.e., has no influence on main topology properties of $\mathcal{G}_{a}$) \cite{qi2012laplacian,baruah2017comparative}. In general, the Laplacian centrality of node $v_{i}$ is determined by the number of walks it participates in $\mathcal{G}_{a}$, i.e., the number of closed walks $\left(v_{i},\ldots,v_{i}\right)$, the number of non-closed walks $\left(v_{i},\ldots\right)$ and $\left(\ldots,v_{i}\right)$ where $v_{i}$ is one of the end nodes, the number of non-closed walks $\left(\ldots,v_{i},\ldots\right)$ containing $v_{i}$ as a middle node. In Ref. \cite{qi2012laplacian}, it is suggested that Eq. (\ref{EQ38}) can be reformulated by analyzing walks of length $2$ (contain $3$ nodes). Specifically, one can derive Eq. (\ref{EQ39}) following Ref. \cite{qi2012laplacian}
\begin{align}
    \mathcal{L}\left(\mathcal{G}_{a}\right)-\mathcal{L}\left(\mathcal{G}_{a}/\{v_{i}\}\right)&=4\eta_{c}\left(v_{i}\right)+2\eta_{e}\left(v_{i}\right)+2\eta_{m}\left(v_{i}\right),\label{EQ39}
\end{align}
where $\eta_{c}\left(v_{i}\right)$ measures the number of closed walks $\left(v_{i},v_{j},v_{i}\right)$, $\eta_{e}\left(v_{i}\right)$ measures the number of non-closed walks $\left(v_{i},v_{j},v_{k}\right)$ and $\left(v_{j},v_{k},v_{i}\right)$, $\eta_{m}\left(v_{i}\right)$ measures the number of non-closed walks $\left(v_{j},v_{i},v_{k}\right)$. Applying the weight adjacent matrices $W\left(a\right)$ and $M\left(a\right)$ (here $M\left(a\right)$ is the weight adjacent matrix of network $\mathcal{G}_{a}/\{v_{i}\}$), we suggest that Eq. (\ref{EQ39}) can be calculated as
\begin{align}
    &\mathcal{L}\left(\mathcal{G}_{a}\right)-\mathcal{L}\left(\mathcal{G}_{a}/\{v_{i}\}\right)\notag\\=&4\left[W\left(a\right)^{2}\right]_{ii}+2\left[\sum_{j\neq k}\left[W\left(a\right)^{2}\right]_{jk}-\sum_{j\neq k}\left[M\left(a\right)^{2}\right]_{jk}\right].\label{EQ40}
\end{align}
Note that $\left[W\left(a\right)^{2}\right]_{jk}=\left[W\left(a\right)W\left(a\right)\right]_{jk}$, the $\left(i,j\right)$ element in matrix $W\left(a\right)^{2}$, is not equivalent to $W_{ij}^{2}\left(a\right)$, the second power of $\left(i,j\right)$ element in matrix $W\left(a\right)$. Combining Eqs. (\ref{EQ37}-\ref{EQ38}) and Eq. (\ref{EQ40}), we can measure the Laplacian centrality of each node in network $\mathcal{G}_{a}$ and only keep $n$ nodes with relatively large Laplacian centrality values. In other words, $m-n$ nodes are filtered because their effects on topology properties of $\mathcal{G}_{a}$ are less significant (see Fig. 5). We refer to the network after filtering as $\widehat{\mathcal{G}}_{a}$ and denote its Laplacian and Gaussian variable as $\widehat{L}\left(a\right)$ and $\widehat{\mathcal{X}}_{\phi}^{a}$, respectively. Then we can approximatively calculate information divergence in Eq. (\ref{EQ20}) by replacing $L\left(a\right)$ and $\mathcal{X}_{\phi}^{a}$ as $\widehat{L}\left(a\right)$ and $\widehat{\mathcal{X}}_{\phi}^{a}$ (see Fig. 5). In the case where $m<n$, we can similarly deal with network $\mathcal{G}_{b}$ following the above approach.

The rationality $\gamma$ of the above approximation can be measured based on the loss of Laplacian energy. Taking the case where $m>n$ as an instance, we define the rationality of approximating $\mathcal{G}_{a}$ by $\widehat{\mathcal{G}}_{a}$ as (see Fig. 5)
\begin{align}
    \gamma\left(\mathcal{L}\left(\mathcal{G}_{a}\right),\mathcal{L}\left(\widehat{\mathcal{G}}_{a}\right)\right)&=\frac{\mathcal{L}\left(\widehat{\mathcal{G}}_{a}\right)}{\mathcal{L}\left(\mathcal{G}_{a}\right)}.\label{EQ41}
\end{align}

To this point, we have presented analytic metrics of network relations from the perspectives of encoding, decoding, and causal analyses in Sec. \ref{Sec4}. We have also explored their generalization in Sec. \ref{Sec5}. Below, we validate our approach on representative complex networks to define encoding, decoding, and causal analysis tasks.

\section{Encoding, decoding, and causal analyses on random network models}\label{Sec6}
We first consider encoding, decoding, and causal analyses on random network models, such as Watts-Strogatz model (with small-world properties) \cite{watts1998collective}, Erdos-Renyi model (each pair of nodes are connected according to a probability quantity) \cite{erdHos1960evolution}, and Barab\'{a}si–Albert model (with scale-free properties) \cite{barabasi1999emergence}. These random network models are important in statistical physics and mathematics (e.g., for analyzing percolation on small-world networks \cite{moore2000epidemics}, Erdos-Renyi networks \cite{almeira2020scaling}, and scale-free networks \cite{xulvi2003correlations}). Meanwhile, they are prototypes in analyzing social \cite{csanyi2004structure,zekri2001statistical}, biological \cite{hallquist2018graph,zhao2021regional,benigni2021persistence,sotero2020estimation,shirai2020long,tian2022percolation}, and chemical \cite{martinez2021computational,emmerich2012diffusion} networks. Therefore, the encoding, decoding, and causal analyses implemented on these models can be further generalized to diverse real networks with corresponding topology properties. The main motivation of our analyses on random network models is to validate the proposed analytic metrics of network relations and suggest practical solutions of potential limitations. 

 \begin{figure*}[t!]
\includegraphics[width=1\columnwidth]{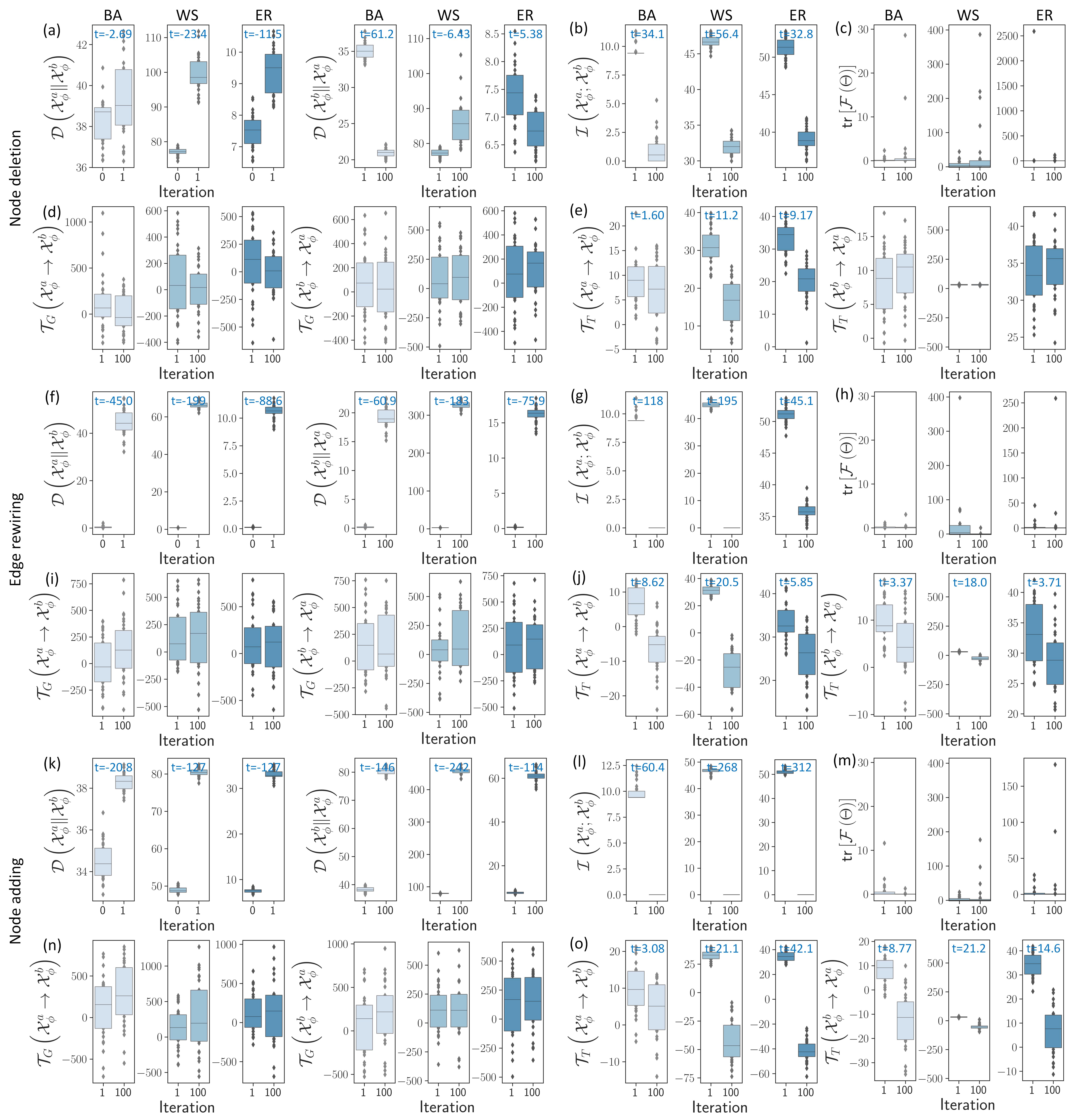}
\caption{\label{G6} Encoding, decoding, and causal analyses on Watts-Strogatz (WS), Erdos-Renyi (ER), and Barab\'{a}si–Albert (BA) random networks during node deletion, edge rewiring, and node adding processes. (a-e) The comparison between the encoding, decoding, and causality metrics derived in the the $1$-th and $100$-th iterations of the node deletion process. (f-j) The comparison between the encoding, decoding, and causality metrics derived in the the $1$-th and $100$-th iterations of the edge rewiring process. (l-p) The comparison between the encoding, decoding, and causality metrics derived in the the $1$-th and $100$-th iterations of the node adding process. Note that the corresponding $t$-statistic will be shown along with the data if the comparison can pass the $t$-test with a standard of $p<10^{-4}$.} 
 \end{figure*}

 \subsection{Experiment designs}
In our experiment, a Watts-Strogatz model (each node originally connects with $\alpha=15$ nodes, and edges are randomly re-wired according a probability of $\beta=0.7$), an Erdos-Renyi model (each pair of nodes are connected according to a probability of $\rho=0.15$), and a Barab\'{a}si–Albert model (there are $50$ edges that bridge between a new node to existing nodes during network construction) are generated and initially contain $300$ nodes. Note that all the network parameters used in initialization are set for convenience, and our analyses do not critically relay on these parameters.

We consider three representative network evolution processes, i.e., node deletion, edge rewiring, and node adding, on these initialized networks, where each process consists of $100$ iterations. During the node deletion process, we randomly delete one node and all related edges from these three networks in each iteration. During the edge rewiring process, we randomly select one node and rewire its edges in each iteration. The edge rewiring rules can be set in diverse forms but should be different from the edge wiring rules in original random networks (e.g., the rewiring rules in the Barab\'{a}si–Albert random network should not be preferential attachment). Otherwise, the generated network after rewiring may not become increasingly different from the original one as the iteration number $n$ increases. To ensure the enlarging difference, we define the rewiring processes of initialized Watts-Strogatz, Erdos-Renyi, and Barab\'{a}si–Albert random networks following the wring rules of Erdos-Renyi ($\rho=0.5$), Watts-Strogatz ($\alpha=40$ and $\beta=0.1$), and Erdos-Renyi ($\rho=0.1$) models, respectively. During the node adding process, we add a node in each iteration and connect it with existing nodes according to certain wiring rules. The wiring rules are set to be distinct from those in the original networks. For convenience, we design the wiring of added nodes in initialized Watts-Strogatz, Erdos-Renyi, and Barab\'{a}si–Albert networks following the rules in Erdos-Renyi ($\rho=0.5$), Watts-Strogatz ($\alpha=40$ and $\beta=0.1$), and Erdos-Renyi ($\rho=0.6$) models, respectively.

Encoding, decoding, and causal analyses are implemented between $\mathcal{X}_{\phi}^{b}$, the networks in the $n$-th iteration ($n\in\mathbb{Z}\cap\left[1,100\right]$) and $\mathcal{X}_{\phi}^{a}$, the networks in their initialized forms. Certainly, one may notice that the decoding analysis has not been explicitly defined by the above settings. In real cases, the decoding analysis should be defined according to research demands. In our research, we present an instance of the decoding analysis based on a parameter vector $\Theta=\left(\theta_{1},\ldots,\theta_{10}\right)$ controlled by $\mathcal{X}_{\phi}^{b}$. Parameter vector $\Theta$ is designed to affect $\mathcal{X}_{\phi}^{a}$ and make it become $\mathcal{X}_{\phi}^{a}+\varepsilon$, where $\varepsilon\sim\mathcal{N}\left(\mathbf{0},\Theta\right)$ denotes the effects of $\Theta$ on $\mathcal{X}_{\phi}^{a}$. For convenience, we consider a case where $\Theta$ is defined as the degree vector $\left(\theta_{1}=\operatorname{deg}\left(v_{1}\right),\ldots,\theta_{10}=\operatorname{deg}\left(v_{10}\right)\right)$ of a set of nodes $\{v_{1},\ldots,v_{10}\}$ randomly selected from network $\mathcal{G}_{b}$. By repeating random sampling, we can obtain a set of observations $\{\Theta^{i}=\left(\theta_{1}^{i},\ldots,\theta_{10}^{i}\right)\}$ of the parameter vector, each of which corresponds to an effect on $\mathcal{X}_{\phi}^{a}$ to create an observation $\mathcal{X}_{\phi}^{a}+\varepsilon_{i}$. Based on these settings, the decoding analysis can be implemented to measure the information of $\Theta$ contained in the probability distribution of $\mathcal{X}_{\phi}^{a}+\varepsilon$. 

In our experiment, each kind of network evolution process is repeated for $50$ times such that encoding, decoding, and causal analyses can be implemented on different realizations of random network evolution.

 \subsection{Experiment results}
 As $n$ increases, more topology properties are changed due to node deletion, edge rewiring, or node adding. Therefore, $\mathcal{X}_{\phi}^{a}$ and $\mathcal{X}_{\phi}^{b}$ are expected to become increasingly different during network evolution processes. Below, we validate whether the enlarging difference can be captured by our encoding, decoding, and causal analyses. 
 
 In Fig. 6, we compare between the results of encoding, decoding, and causal analyses obtained in the $1$-th and $100$-th iterations of network evolution processes. The changes of these relation metrics are statistically significant if they can pass the $t$-test \cite{gerald2018brief} (i.e., the distributions of these metrics in the $1$-th and $100$-th iterations are statistically different). Otherwise, they should be treated as less sensitive to network evolution. As shown in the experiment results, information divergence, mutual information, and conditional mutual information can robustly pass the $t$-test with a rigid standard of $p<10^{-4}$. As presented in Fig. 6, these statistically significant relation metrics can reflect the reduction of network similarities during network evolution. Specifically, the enlarging difference between $\mathcal{X}_{\phi}^{a}$ and $\mathcal{X}_{\phi}^{b}$ can be generally reflected by the increasing information divergence, the decreasing mutual information, and the decreasing conditional mutual information. An exception to this property is the decreasing information divergence from $\mathcal{X}_{\phi}^{b}$ to $\mathcal{X}_{\phi}^{a}$ during node deletion. We hypothesize that the inconsistent variation trends of $\mathcal{D}\left(\mathcal{X}_{\phi}^{a}\Vert\mathcal{X}_{\phi}^{b}\right)$ and $\mathcal{D}\left(\mathcal{X}_{\phi}^{b}\Vert\mathcal{X}_{\phi}^{a}\right)$ during node deletion may arise from the asymmetry properties of information divergence (i.e., information divergence is a kind of pseudo-distance). Meanwhile, the information divergence generalized by the Laplacian-energy-based approach in Sec. \ref{Sec5} may fail to reflect the actual divergence between networks with different sizes. As for the other metrics that are not statistically significant in the $t$-test (e.g., Fisher information and Granger causality), they do not have clear patterns at the group level and exhibit high diversities across different realizations of network evolution processes. This phenomenon may arise from the numerical susceptibility of these metrics towards concrete configurations of random networks. 
 
 In sum, we suggest information divergence, mutual information, and conditional mutual information as prior choices for analyzing random network evolution. The results derived by Fisher information and Granger causality may be more numerically susceptible to the topology properties of concrete random network realizations.

 \section{Encoding, decoding, and causal analyses on real networks}\label{Sec7}
 Among various tasks in network analysis, assigning network similarity in network ensembles is an important one, which is closely related to network clustering, query, and classification tasks in machine learning. Here we implement the similarity measurement task under our theoretical framework and other competitive alternatives. 

 \subsection{Data sets}
Three real network data sets are used in our experiment. The first data set, PROTEINS \cite{borgwardt2005protein}, contains the network structures of numerous proteins. These proteins are classified into enzymes and non-enzymes classes. The second data set, MUTAG \cite{debnath1991structure}, is a collection of mutagenic aromatic and heteroaromatic nitro compounds. Each chemical compound is represented by a network of atoms and is classified into two classes according to their mutagenic effects on specific gram negative bacteria. The third data set, ENZYMES \cite{borgwardt2005protein}, contains protein tertiary structures (i.e., the structure where polypeptide chains become functional) derived from the BRENDA enzyme data. There are six kinds of enzymes included in the data set. These three data sets are filtered such that all remaining networks are connected graphs (i.e., each network has one connected component). Meanwhile, PROTEINS and ENZYMES data sets are filtered according to network size to ensure that remaining networks are not too small (i.e., thresholds are set as $50$ and $40$ for PROTEINS and ENZYMES, respectively). Note that the filtering procedure is proposed for numerical convenience as some of the compared approaches in our experiment may meet numerical issues on small or disconnected networks.

 \begin{figure*}[t!]
\includegraphics[width=1\columnwidth]{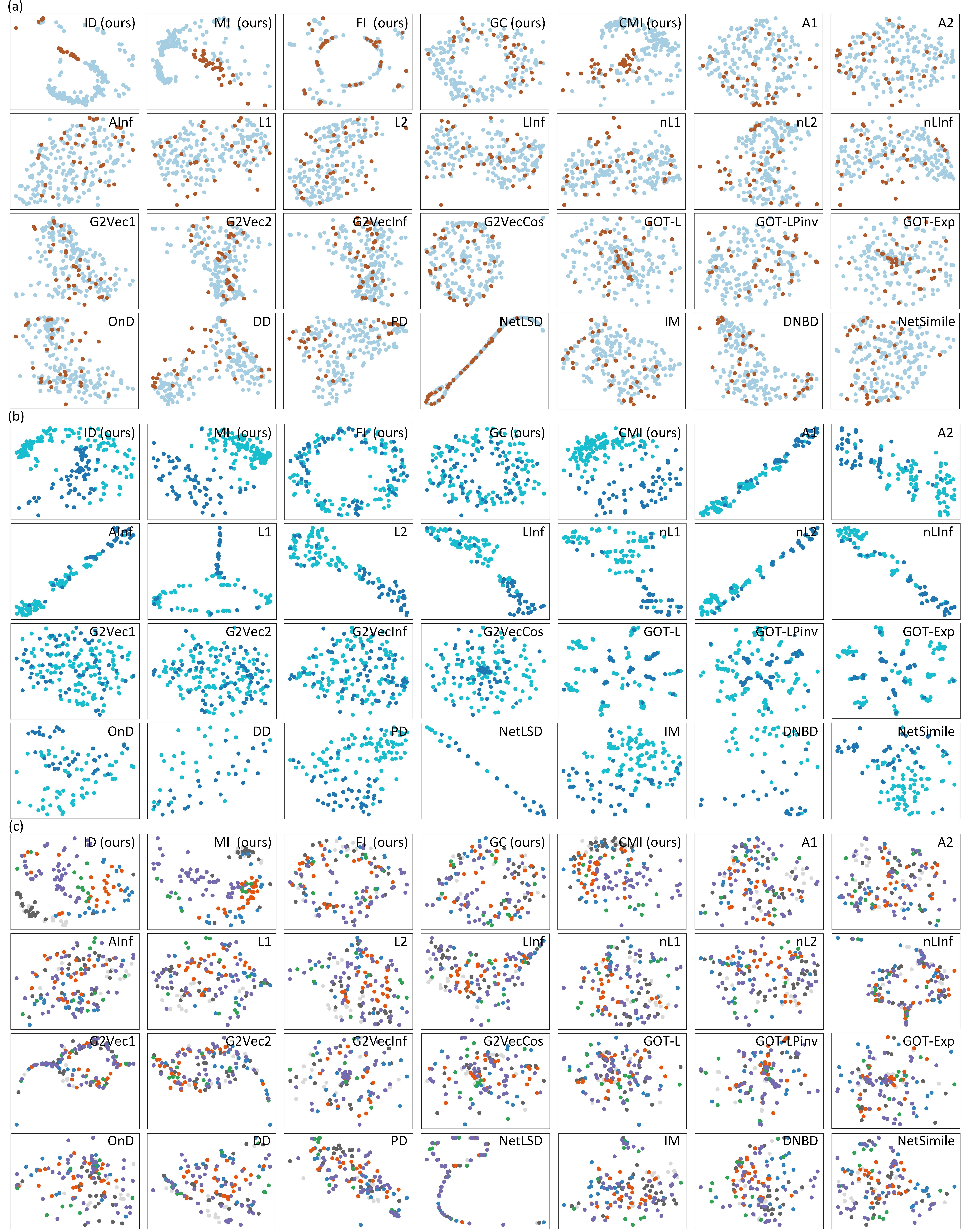}
\caption{\label{G7} The $t$-SNE constrained by different network relations on (a) PROTEINS, (b) MUTAG, and (c) ENZYMES data sets. } 
 \end{figure*}

 \begin{figure*}[t!]
\includegraphics[width=1\columnwidth]{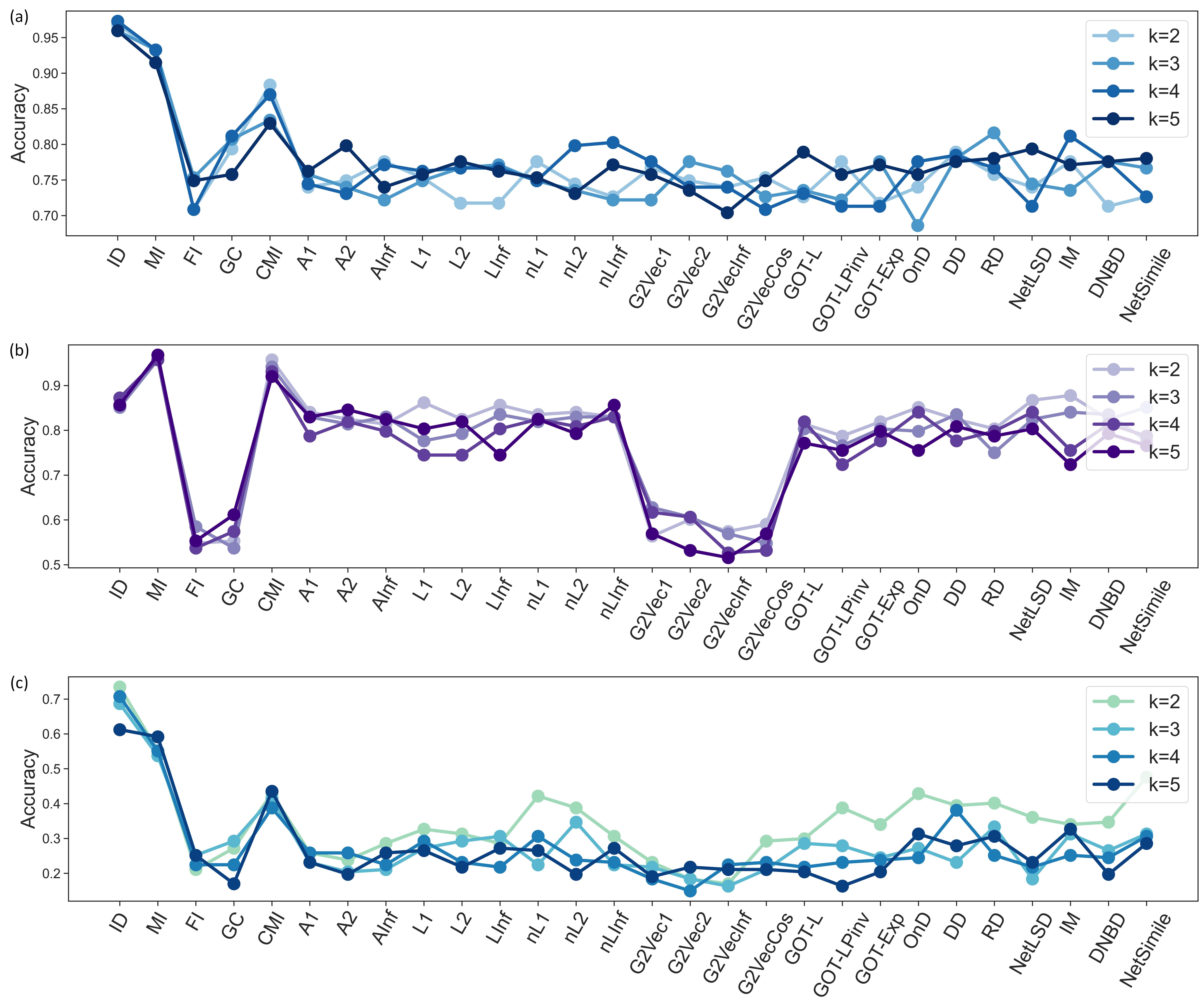}
\caption{\label{G8} The $k$-nearest neighbor query accuracy ($k\in\{2,3,4,5\}$) evaluated on the $t$-SNE constrained by different network relations on (a) PROTEINS, (b) MUTAG, and (c) ENZYMES data sets. } 
 \end{figure*}

\subsection{Compared approaches}
Apart from our proposed information divergence (ID), mutual information (MI), Fisher information (FI), Granger causality (GC), and conditional mutual information (CMI), numerous classic approaches are also implemented in our experiment for comparison. 

The first family of approaches, provided by the NetComp toolbox \cite{wills2020metrics},  assign the difference between two networks by calculating the $\mathsf{L}_{p}$ distance between the $m$ largest eigenvalues of adjacency matrices (A), Laplacian operators (L), and normalized Laplacian operators (nL). For each matrix representation, we calculate $\mathsf{L}_{1}$, $\mathsf{L}_{2}$, and $\mathsf{L}_{\infty}$ distances to measure similarities (we set $m=10$ for convenience). According to the selection of distance and matrix representation, the results are referred to as A1, A2, AInf, L1, L2, LInf, nL1, nL2, and nLInf, respectively. For instance, nLInf refers to the $\mathsf{L}_{\infty}$ distance between the eigenvalue vectors of normalized Laplacian operators. Other metrics can be interpreted in similar ways.

The second family of approaches embed networks as vectors applying the Graph2Vec framework \cite{narayanan2017graph2vec} (note that we set the embedded vector length as $128$). Network difference is measured as the $\mathsf{L}_{p}$ or cosine distance between the embedded vectors of networks. According to the applied distance, the derived results are referred to as G2Vec1, G2Vec2, G2VecInf, and G2VecCos, respectively. For example, G2Vec1 refers to the $\mathsf{L}_{1}$ distance while G2VecCos denotes the cosine distance.

The third family of approaches are rooted in the theory of optimal transport between graphs \cite{petric2019got,maretic2022fgot}. In general, we represent each network $\mathcal{G}$ as a Gaussian variable $\mathcal{X}_{\phi}\sim\mathcal{N}\left(\mathbf{0},f\left(L\right)\right)$, where $f\left(L\right)$ denotes a function of Laplacian $L$. In our experiment, we set $f\left(L\right)=L+\frac{1}{n}J$ (same as Eq. (\ref{EQ16}) and referred to as GOT-L), $f\left(L\right)=L^{\dagger}+\frac{1}{n}J$ (referred to as GOT-LPinv), and $f\left(L\right)=\exp\left(-\tau L\right)$ (we set $\tau=0.5$ and refer to it as GOT-Exp). Then, we can analytically derive the Wasserstein distance between each pair of the defined Gaussian variables to assign the difference between corresponding networks. Please see Refs. \cite{petric2019got,maretic2022fgot} for detailed calculation approaches.

Other considered approaches are proposed from diverse perspectives and have distinct characteristics (see Ref. \cite{hartle2020network} for a comprehensive review). In our experiment, we apply onion divergence (OnD, the Jensen-Shannon divergence between the onion decomposition results of two networks) \cite{hebert2016multi,allard2019percolation}, degree divergence (DD, the Jensen-Shannon divergence measured between degree distributions) \cite{carpi2011analyzing}, portrait divergence (PD, the Jensen-Shannon divergence between network
portraits) \cite{bagrow2019information}, NetLSD distance (NetLSD, the Frobenius norm of the difference between the heat trace signatures of normalized Laplacian operators) \cite{tsitsulin2018netlsd}, Ipsen-Mikhailov distance (IM, a kind of spectral comparison between Laplacian operators) \cite{ipsen2004evolutionary}, distributional non-backtracking spectral distance (DNBD, the difference between the eigenvalues of the non-backtracking matrices of networks) \cite{torres2019non}, and NetSimile (NetSimile, the difference between multiple statistical features of networks) \cite{berlingerio2012netsimile}.

\subsection{Experiment designs}

Our experiment consists of three main steps. First, we calculate each network relation metric among networks to generate the corresponding network relation matrix $R$, where $R_{ij}$ denotes the relation metric between networks $\mathcal{G}_{i}$ and $\mathcal{G}_{j}$ (e.g., when mutual information is considered, element $R_{ij}$ measures the mutual information between $\mathcal{G}_{i}$ and $\mathcal{G}_{j}$). For the network relation metrics whose larger values suggest larger differences between networks (e.g., our proposed information divergence and all the implemented classic relation metrics), matrix $R$ can directly serve as a distance matrix, denoted by $D$. For the network relation metrics whose larger values denote larger similarities between networks (e.g., our proposed mutual information, Fisher information, Granger causality, and conditional mutual information), we transform matrix $R$ to a distance matrix $D$ following $D_{ij}=-R_{ij}+\max_{i,j}R_{ij}$. For the sake of simplicity, we average between $D_{ij}$ and $D_{ji}$ to make the derived distance matrix symmetric.

Second, we use these distance matrices to constrain the computing process of the $t$-SNE analysis \cite{van2008visualizing}, a kind of unsupervised machine learning approach for dimension reduction. The constraint is realized by replacing the default distance measurement in the $t$-SNE analysis (e.g., $\mathsf{L}_{p}$ distance in most common toolboxes) with our pre-computed distance matrix $D$. Based on this setting, the results of the $t$-SNE can reflect the properties of the customized matrix $D$. In our experiment, we apply the $t$-SNE to embed networks into a two dimensional space.

Third, we evaluate the validity of the $t$-SNE constrained by matrix $D$ by a $k$-nearest neighbor query task. In this task, we search the $k$-nearest neighbor for each network in the embedded space and compare between their labels (i.e., the class information). The $k$-nearest neighbor is determined using the $\mathsf{L}_{2}$ distance. If the metric used for defining matrix $D$ can properly capture the relation between networks, the selected network is expected to share the same label with its $k$-nearest neighbor (i.e., they belong to the same class) when $k$ is not too large. We treat a query as correct if the network and its queried neighbor belong to the same class. Otherwise, the query is treated as wrong. By implementing queries on all networks in the embedded space, we can calculate the query accuracy (i.e., the proportion of correct queries among all queries) to evaluate the validity of network relation metric. An ideal network relation metric should achieve a high query accuracy on each data set. 

\subsection{Experiment results}

For each data set, we visualize its embedded spaces derived from the $t$-SNE analysis constrained by different network relation metrics in Fig. 7. Class labels are distinguished according to colors. Compared with classic approaches, our information divergence (ID), mutual information (MI), and conditional mutual information (CMI) create more clear data distributions in the embedded spaces, where each class of networks are close to each other to form a cluster with clear patterns. As for Fisher information (FI), Granger causality (GC), and classic approaches, they imply more blurry embedded distributions with low data separability between different classes.

To quantitatively validate our above observations, we present the $k$-nearest neighbor query accuracy associated with each network relation metric ($k\in\{2,3,4,5\}$) in Fig. 8. Consistent with Fig. 7, the  query accuracy values achieved by ID, MI, and CMI are higher than those achieved by other metrics. FI and GC generally achieve similar accuracy values with classic approaches. These results suggest the applicability of our framework in characterizing network relations.

In sum, we have compared our proposed network relation metrics with other approaches \cite{wills2020metrics,narayanan2017graph2vec,petric2019got,maretic2022fgot,hartle2020network,hebert2016multi,allard2019percolation,carpi2011analyzing,bagrow2019information,tsitsulin2018netlsd,ipsen2004evolutionary,torres2019non,berlingerio2012netsimile} in network embedding and query tasks. Our experiments demonstrate that our framework can achieve competitive or better performance in these tasks, suggesting the potential of our approach in studying diverse science and engineering questions related to network comparison.

\section{Discussion}\label{Sec8}
\subsection{Progress compared with previous works}\label{Sec8-1}
Compared with previous data-driven works \cite{soundarajan2014guide,attar2017classification,conte2004thirty,foggia2014graph,yan2016short,gao2010survey,borgwardt2020graph}, one of the main progress accomplished in our research is to suggest a general way to represent the topology properties of an arbitrary network. Our theory explores a mapping $\phi$ to map an arbitrary network $\mathcal{G}\left(V,E\right)$ to a random variable $\mathcal{X}_{\phi}$ distributed on node set $V$. The random variable is defined as $\mathcal{X}_{\phi}\sim\mathcal{N}\left(\mathbf{0},L+\frac{1}{n}J\right)$, a Gaussian variable characterized by a function of the Laplacian $L$ of network $\mathcal{G}$. On the one hand, such a definition ensures that the average smoothness of mapping $\phi$ on network $\mathcal{G}$ is fully determined by the information of network topology properties contained in Laplacian $L$. On the other hand, this definition satisfies the requirements of maximum entropy property of variable $\mathcal{X}_{\phi}$ to promote the applicability in measuring information quantities. Based on $\mathcal{X}_{\phi}$, we further define encoding (information divergence and mutual information), decoding (Fisher information), and causal analyses (Granger causality and conditional mutual information) between complex networks. We have validated these analyses on random network models, protein-protein interaction network, and chemical compound network ensembles. Our proposed metrics, especially information divergence, mutual information, and conditional mutual information, can properly capture the dynamic evolution of random networks and outperform classic approaches \cite{wills2020metrics,narayanan2017graph2vec,petric2019got,maretic2022fgot,hartle2020network,hebert2016multi,allard2019percolation,carpi2011analyzing,bagrow2019information,tsitsulin2018netlsd,ipsen2004evolutionary,torres2019non,berlingerio2012netsimile} in the comparison between real networks. A release of our algorithms can be found in Ref. \cite{yang2022toolbox}. In the future, one can further consider Fisher-Rao metric in information geometry \cite{ning2018smooth,ning2019smooth} and Wasserstein-2 metric in optimal transport \cite{ning2018smooth,ning2019smooth,petric2019got,dong2020copt}, both of which can be readily generalized to Gaussian variables.

\subsection{Mathematical relations between our theory and related results}\label{Sec8-2}

To understand the difference between our work and previous studies \cite{dong2016learning,kalofolias2016learn,petric2019got,dong2020copt}, we begin with discussing the meaning of the following covariance matrix
 \begin{align}
    \Sigma^{\heartsuit}=L^{\dagger}+\frac{1}{n}J,\label{EQ43}
\end{align}
which is directly related to the covariance matrix in Refs. \cite{dong2016learning,kalofolias2016learn,petric2019got,dong2020copt}. As suggested by Eq. (\ref{EQ9}), the covariance matrix in Eq. (\ref{EQ43}) is exactly the inverse of our result $\Sigma=L+\frac{1}{n}J$, i.e., $\Sigma^{\heartsuit}=\Sigma^{-1}$. If one defines a Gaussian variable $\mathcal{X}_{\phi}^{\heartsuit}\sim\mathcal{N}\left(\mathbf{0},\Sigma^{\heartsuit}\right)$, then its precision matrix $Q^{\heartsuit}$ equals our proposed covariance matrix
 \begin{align}
    Q^{\heartsuit}:={\Sigma^{\heartsuit}}^{-1}=\Sigma.\label{EQ44}
\end{align}
Because the partial correlation between $X_{\phi}^{\heartsuit}\left(i\right)$ and $X_{\phi}^{\heartsuit}\left(j\right)$, the actual values of $\mathcal{X}_{\phi}^{\heartsuit}$ on nodes $v_{i}$ and $v_{j}$, is fully characterized by the precision matrix \cite{rue2005gaussian}
 \begin{align}
    &\operatorname{corr}\left(X_{\phi}^{\heartsuit}\left(i\right),X_{\phi}^{\heartsuit}\left(j\right)\big\vert\mathcal{X}_{\phi}^{\heartsuit}\setminus\{X_{\phi}^{\heartsuit}\left(i\right),X_{\phi}^{\heartsuit}\left(j\right)\}\right)\notag\\:=&-\frac{Q^{\heartsuit}_{ij}}{\sqrt{Q^{\heartsuit}_{ii}Q^{\heartsuit}_{jj}}},\label{EQ45}\\=&-\frac{\Sigma_{ij}}{\sqrt{\Sigma_{ii}\Sigma_{jj}}},\label{EQ46}
\end{align}
it is trivial to know that variables $X_{\phi}^{\heartsuit}\left(i\right)$ and $X_{\phi}^{\heartsuit}\left(j\right)$ are expected to have a stronger partial correlation if nodes $v_{i}$ and $v_{j}$ are connected by an edge with larger weight (i.e., a larger value of $-\Sigma_{ij}$). Similarly, variables $X_{\phi}^{\heartsuit}\left(i\right)$ and $X_{\phi}^{\heartsuit}\left(j\right)$ are expected to share no significant relation if nodes $v_{i}$ and $v_{j}$ are dis-connected. Therefore, the Gaussian variable defined by $\Sigma^{\heartsuit}$ is more applicable to the cases where edge weights reflect the consistent relations between nodes (i.e., positive correlations or coherence). According to Eq. (\ref{EQ10}). the expected smoothness of mapping $\phi$ in such a Gaussian variable is
 \begin{align}
     \mathbb{E}\left(\mathcal{S}\left(\phi\right)\right)&=\mathbb{E}\left(\mathcal{X}_{\phi}\right)^{T}L\mathbb{E}\left(\mathcal{X}_{\phi}\right)+\operatorname{tr}\left(L\Sigma^{\heartsuit}\right),\label{EQ47}\\
     &=\operatorname{tr}\left[L\left(L^{\dagger}+\frac{1}{n}J\right)\right],\label{EQ48}\\
     &=\operatorname{tr}\left(I\right)-\frac{1}{n}\operatorname{tr}\left(J\right)+\operatorname{tr}\left[L\left(I-L^{\dagger}L\right)\right],\label{EQ49}\\
     &=n-1.\label{EQ50}
\end{align}
 Eqs. (\ref{EQ49}-\ref{EQ50}) are derived from the fact that $LL^{\dagger}L=L$ \cite{barata2012moore} and $LL^{\dagger}=L^{\dagger}L=I-\frac{1}{n}J$ \cite{gutman2004generalized,chebotarev2006proximity,van2017pseudoinverse}. In general, Eqs. (\ref{EQ47}-\ref{EQ50}) mean that the expected smoothness of mapping $\phi$ in a network characterized by $\mathcal{X}_{\phi}^{\heartsuit}\sim\mathcal{N}\left(\mathbf{0},\Sigma^{\heartsuit}\right)$ is independent of the network topology properties conveyed by $L$. On the contrary, the expected smoothness is fully determined by the network size. We speculate that this property may limit the capacity of $\mathcal{X}_{\phi}^{\heartsuit}\sim\mathcal{N}\left(\mathbf{0},\Sigma^{\heartsuit}\right)$ to describe complex networks with high heterogeneity.
 
In our results, the Gaussian variable is $\mathcal{X}_{\phi}\sim\mathcal{N}\left(\mathbf{0},L+\frac{1}{n}J\right)$, where the expected smoothness of mapping $\phi$ is fully characterized by $L$ via $\mathbb{E}\left(\mathcal{S}\left(\phi\right)\right)=\operatorname{tr}\left(LL\right)$ (see Eq. (\ref{EQ14})). The covariance matrix $\Sigma$ of such a random variable implies that variables $X_{\phi}\left(i\right)$ and $X_{\phi}\left(j\right)$ are expected to evolve inversely (i.e., stronger negative covariance) if nodes $v_{i}$ and $v_{j}$ are connected by an edge with larger weight. Meanwhile, variables $X_{\phi}\left(i\right)$ and $X_{\phi}\left(j\right)$ have no significant relation if nodes $v_{i}$ and $v_{j}$ share no edge between them. In other words, the Gaussian variable defined by $\Sigma$ is more applicable to the cases where edge weights reflect opposite relations between nodes (i.e., negative correlations or anti-coherence). Meanwhile, it may have higher potential to characterize heterogeneous networks where node difference matters. 

In sum, the proposed covariance matrix $\Sigma=L+\frac{1}{n}J$ and its inverse $\Sigma^{\heartsuit}=L^{\dagger}+\frac{1}{n}J$ are applicable to opposite conditions, respectively. Although we primarily use $\Sigma$ to define encoding, decoding, and causal analyses in our paper, all derived results can be readily reformulated using $\Sigma^{\heartsuit}$. Our released toolbox \cite{yang2022toolbox} allows users to choose between $\Sigma$ and $\Sigma^{\heartsuit}$ for a better network characterization.  

\subsection{Limitations}\label{Sec8-3}
As an initial attempt, there remain limitations in our work for further exploration. Here we suggest two limitations whose solutions may advance related fields. 

The first limitation arises from the requirements of non-negative edge weights in defining the Laplacian $L$. There exist numerous real networks whose edge weights can be negative (e.g., in neural populations, inhibitory synapses have negative weights). Although the effects of negative weights on the eigenvalues of $L$ have drawn increasing attention (e.g., see Refs. \cite{ahmadizadeh2017eigenvalues,bronski2016graph,bronski2015spectral}), an optimal definition of $L$ on networks with negative weights remains exclusive. Similarly, the second limitation occurs when one considers networks with directed edges. While notable progress has been accomplished in defining $L$ on directed networks \cite{bauer2012normalized,agaev2005spectra}, these achievements can not completely address our problems because an asymmetric version of $L$ do not support the definition of Gaussian variable (the covariance matrix $\Sigma$ must be symmetric).

\section*{Acknowledgments}
Author Y.T. conceptualizes the idea, develops theoretical frameworks, designs computational tools, and writes the manuscript. Authors H.D.H. and G.Z.X contribute equally to literature collection, mathematics proofreading, and manuscript revision. Author Z.Y.Z. contributes to technical support. Author P.S. contributes to idea conceptualization, manuscript writing, and project supervision. Authors are grateful for the kind helps from Aohua Cheng, a member of the Tsien Excellence in Engineering
Program at Tsinghua University, and Moufan Li, a student from the Department of Computer Science at Tsinghua University. This project is supported by the Artificial and General Intelligence Research Program of Guo Qiang Research Institute at Tsinghua University (2020GQG1017), the Huawei Innovation Research Program (TC20221109044), and the Tsinghua University Initiative Scientific Research Program.


\bibliography{apssamp}
\end{document}